\documentstyle{mn}
\input{epsf.tex}
\newif\ifAMStwofonts

\ifoldfss
  \ifCUPmtlplainloaded \else
    \NewTextAlphabet{textbfit} {cmbxti10} {}
    \NewTextAlphabet{textbfss} {cmssbx10} {}
    \NewMathAlphabet{mathbfit} {cmbxti10} {} 
    \NewMathAlphabet{mathbfss} {cmssbx10} {} 
  \fi
  \ifAMStwofonts
    \ifCUPmtlplainloaded \else
      \NewSymbolFont{upmath} {eurm10}
      \NewSymbolFont{AMSa} {msam10}
      \NewMathSymbol{\upi}     {0}{upmath}{19}
      \NewMathSymbol{\umu}     {0}{upmath}{16}
      \NewMathSymbol{\upartial}{0}{upmath}{40}
      \NewMathSymbol{\leqslant}{3}{AMSa}{36}
      \NewMathSymbol{\geqslant}{3}{AMSa}{3E}

       \let\le=\leqslant
       \let\ge=\geqslant
    \fi
  \fi
\fi 

\ifnfssone
  \newmathalphabet{\mathit}
  \addtoversion{normal}{\mathit}{cmr}{m}{it}
  \addtoversion{bold}{\mathit}{cmr}{bx}{it}
  \newmathalphabet{\mathbfit} 
  \addtoversion{normal}{\mathbfit}{cmr}{bx}{it}
  \addtoversion{bold}{\mathbfit}{cmr}{bx}{it}
  \newmathalphabet{\mathbfss} 
  \addtoversion{normal}{\mathbfss}{cmss}{bx}{n}
  \addtoversion{bold}{\mathbfss}{cmss}{bx}{n}
  \ifAMStwofonts
    \ifCUPmtlplainloaded \else
      \UseAMStwoboldmath
      \makeatletter
      \new@mathgroup\upmath@group
      \define@mathgroup\mv@normal\upmath@group{eur}{m}{n}
      \define@mathgroup\mv@bold\upmath@group{eur}{b}{n}
      \edef\UPM{\hexnumber\upmath@group}
      \new@mathgroup\amsa@group
      \define@mathgroup\mv@normal\amsa@group{msa}{m}{n}
      \define@mathgroup\mv@bold\amsa@group{msa}{m}{n}
      \edef\AMSa{\hexnumber\amsa@group}
      \makeatother
      \mathchardef\upi="0\UPM19
      \mathchardef\umu="0\UPM16
      \mathchardef\upartial="0\UPM40
      \mathchardef\leqslant="3\AMSa36
      \mathchardef\geqslant="3\AMSa3E

       \let\le=\leqslant
       \let\ge=\geqslant
    \fi
  \fi
\fi 

\ifnfsstwo
  \DeclareMathAlphabet{\mathbfit}{OT1}{cmr}{bx}{it}
  \SetMathAlphabet\mathbfit{bold}{OT1}{cmr}{bx}{it}
  \DeclareMathAlphabet{\mathbfss}{OT1}{cmss}{bx}{n}
  \SetMathAlphabet\mathbfss{bold}{OT1}{cmss}{bx}{n}
  \ifAMStwofonts
    \ifCUPmtlplainloaded \else
      \DeclareSymbolFont{UPM}{U}{eur}{m}{n}
      \SetSymbolFont{UPM}{bold}{U}{eur}{b}{n}
      \DeclareSymbolFont{AMSa}{U}{msa}{m}{n}
      \DeclareMathSymbol{\upi}{0}{UPM}{"19}
      \DeclareMathSymbol{\umu}{0}{UPM}{"16}
      \DeclareMathSymbol{\upartial}{0}{UPM}{"40}
      \DeclareMathSymbol{\leqslant}{3}{AMSa}{"36}
      \DeclareMathSymbol{\geqslant}{3}{AMSa}{"3E}

       \let\le=\leqslant
       \let\ge=\geqslant
    \fi
  \fi
\fi 

\ifCUPmtlplainloaded \else
  \ifAMStwofonts \else 
    \def\upi{\pi}
    \def\umu{\mu}
    \def\upartial{\partial}
  \fi
\fi
\def\pn{\par\noindent}
\title{Phase space transport in cuspy triaxial potentials: Can they be used 
to construct self-consistent equilibria?}

\author[C. Siopis and H. E. Kandrup]
  {Christos Siopis$^{1,2}$\thanks{E-mail: siopis@astro.ufl.edu}
    and Henry E. Kandrup,$^{1,2,3,4}$\thanks{E-mail: kandrup@astro.ufl.edu}\\
   $^{1}$ Department of Astronomy, University of Florida, Gainesville, 
          FL 32611, USA\\
   $^{2}$ Observatoire de Marseille, 2 place le Verrier, 13248 Marseille cedex
          04, FRANCE\\
   $^{3}$ Department of Physics, University of Florida, Gainesville, 
          FL 32611, USA\\
   $^{4}$ Institute for Fundamental Theory, University of Florida, Gainesville,
          FL 32611, USA}

\date{Accepted 2000 \hskip 1in .
      Received 2000 \hskip 1in .
      in original form 2000}

\pubyear{1999}

\def\pn{\par\noindent}

\begin{document}

\maketitle

\label{firstpage}

\begin{abstract}
This paper focuses on the statistical properties of chaotic orbit ensembles
evolved in triaxial generalisations of the Dehnen potential which have been
proposed recently to model realistic ellipticals that have a 
strong density cusp and manifest significant deviations from axisymmetry. 
Allowance is made for a possible supermassive black hole, as well as 
low amplitude friction, noise, and periodic driving which can
mimic irregularities associated with discreteness effects and/or an external
environment. The chaos exhibited by these potentials is quantified 
by determining (1) how the relative number of chaotic orbits 
depends on the steepness of the cusp, as probed by ${\gamma}$, the power law
exponent with which density diverges, and $M_{BH}$, the black hole mass; (2) 
how the size of the largest Lyapunov exponent
varies with ${\gamma}$ and $M_{BH}$; and (3) the extent to which
Arnold webs significantly impede phase space transport, both with and without
perturbations. The most important conclusions dynamically are (1) that, in the 
absence of irregularities, chaotic orbits tend to be {\it extremely}
`sticky,' so that different pieces of the same chaotic orbit can behave very
differently for times ${\sim}{\;}10000t_{D}$ or more, but (2) that even very 
low amplitude perturbations can prove efficient in erasing many -- albeit
not all -- of these differences. The 
implications of these facts are discussed both for the structure and evolution 
of real galaxies and for the possibility of constructing approximate 
near-equilibrium models using Schwarzschild's method. For example, when 
trying to use Schwarzschild's method to construct model galaxies containing 
significant numbers of chaotic orbits, it seems advantageous to build 
libraries with chaotic orbits evolved in the presence of low amplitude 
friction and noise, since such noisy orbits are more likely to represent 
reasonable approximations to time-independent building blocks.
Much of the observed qualitative behaviour can be reproduced with a toy 
potential given as the sum of an anisotropic harmonic oscillator and a 
spherical Plummer potential, which suggests that the results may be generic.
\end{abstract}


\section{MOTIVATION}

The work described here exploits recently developed ideas from chaos and 
nonlinear dynamics to better understand the dynamics of some seemingly {\it 
realistic} galactic potentials. These potentials reflect the fact that 
many/most early-type galaxies have a pronounced central density cusp (cf. 
Lauer {\it et al.} 1995), possibly associated with the presence of a 
supermassive black hole (cf. Kormendy \& Richstone 1995); and that, at least 
for galaxies with comparatively shallow cusps, there is often evidence for 
moderate deviations
from axisymmetry (cf. Kormendy \& Bender 1996). This work also embraces the 
fact that real galaxies are continually subjected to various irregularities 
which the theorist might like to ignore, including `high frequency' 
discreteness effects reflecting the existence of internal substructures and 
`lower frequency' effects reflecting, e.g., systematic pulsations or the 
effects of nearby objects. 

Recent interest in chaos in galactic dynamics has been driven primarily by 
data, both ground-based and from the Hubble Space Telescope ({\it HST}), which 
reveal that the density of stars in early-type galaxies typically rises 
towards the center in a power-law cusp (Lauer {\it et al.} 1995, Byun {\it et 
al.} 1996, Gebhardt {\it et al.} 1996, Kormendy {\it et al.} 1996, Moller, 
Stiavelli, \& Zeilinger 1995). For example, an analysis of more
than 65 elliptical and {\it S0} galaxies has established that, at a resolution 
of $<0.1$ arc-seconds, the surface brightness profile $I(R)$ is best 
approximated by a power law profile ${\propto}{\;}R^{-\gamma}$ with $\gamma$ 
ranging from near zero to unity. 
Most previous dynamical studies of galaxies (e.g., the King models) assumed a 
constant density core, with a concomitant analytic central surface brightness, 
$I(R)\propto 1-AR^2+\cdots$. The {\it HST} observations require completely new 
dynamical models to predict kinematic properties of the central regions, and 
to ascertain whether supermassive black holes are actually present. 

There is also evidence that many 
galaxies may be more irregularly shaped than the nearly axisymmetric objects 
assumed as late as the 1970's. For example, twisted isophotes are
interpreted as evidence for deviations from axisymmetry, and the existence 
of nontrivial residuals in a fit of the surface brightness distribution to a 
$\cos 2{\theta}$ law suggests further that many systems do not even exhibit 
the symmetries of a triaxial ellipsoid (cf. Bender {\it et al.} 1989). Indeed,
these observed irregularities are so pronounced that they have been proposed 
as the basis of a new classification scheme for ellipticals (Kormendy and 
Bender 1996). 

The crucial point, then, as stressed by Merritt and collaborators (cf. Merritt
1996, Merritt \& Fridman 1996), is that the combination of cusps and 
triaxiality
seems to make chaos nearly unavoidable. In a non-cuspy triaxial galaxy, the 
central regions are dominated by regular box orbits with the 
topology of a three-dimensional Lissajous figure. Inserting a cusp or a 
supermassive black hole can destabilize these orbits. One thus anticipates 
that many of the orbits passing close to the center of the galaxy must be 
chaotic, and that this feature could play an important role in the structure 
and evolution of these regions of the galaxy. Far from being something exotic 
and improbable, the vast 
majority of elliptical galaxies may contain large fractions of chaotic orbits.

These observations run counter to the historical trend in galactic dynamics, 
the foundations of which go back over fifty years to a time when most 
astronomers had a physical worldview that was dominated by integrable and
near-integrable systems. In recent years, much attention has focused on 
constructing self-consistent galactic models, idealized as time-independent 
solutions to the collisionless Boltzmann equation. In particular, given 
various simplifying assumptions about equilibrium shapes, one can construct 
exact analytic solutions such as the integrable St\"ackel (cf. de Zeeuw 1985) 
models. However, the new {\it HST} observations imply that the very idea of 
integrable self-consistent dynamical models must be rethought (cf. Merritt 
1996). Indeed, as Gebhardt {\it et al.} (1996) put it, `it seems very unlikely 
that experience gained from the analysis of orbits in static St\"ackel 
potentials or of triaxial objects with analytic cores has much connection to 
the central regions of real galaxies.'

Once it be admitted that stellar orbits in galaxies can have a large
chaotic component, a host of fundamental questions arise that to date
have had no fully satisfactory answers: How should one construct
self-consistent equilibria with chaotic orbits? Are the fundamental 
time scales such as the Chandrasekhar relaxation time $t_{R}$ (cf. 
Chandrasekhar 1943a) changed by chaos? 
On what time scale are the trapped chaotic orbits used (cf. Athanassoula
{\it et al.} 1983, Wozniak 1993) to explain the shapes of certain galaxies 
unstable? What is the effect of a large central point mass? How
accurately can the invariant density associated with the chaotic part
of the phase space be approximated? Will elliptical galaxies with
cusps reach triaxial steady states or bypass them in favour of
axisymmetric ones? These, and many other, questions are
variations on the central theme: what are the dynamical consequences
of chaos in galaxies?

The objective here is to explore these issues for the 
triaxial generalisations of the Dehnen (1993) potentials, which have been 
considered extensively by Merritt and collaborators (e.g., Merritt \& Fridman 
1996). These correspond to potentials generated 
self-consistently from the triaxial mass density
\begin{equation}
{\rho}(m)={(3-{\gamma})\over 4{\pi}abc}m^{-{\gamma}}(1+m)^{-(4-{\gamma})}
\end{equation}
with
\begin{equation}
m^{2}={x^{2}\over a^{2}}+{y^{2}\over b^{2}}+{z^{2}\over c^{2}},
\end{equation}
for $c/a=1/2$ and $(a^{2}-b^{2})/(a^{2}-c^{2})=1/2$.
Four cases will be discussed in detail, namely ${\gamma}=0$, ${\gamma}=0.3$,  
${\gamma}=1$, and ${\gamma}=2$, ranging from no cusp to a very steep cusp,
respectively. The 
analysis involves extracting the statistical properties of chaotic orbit 
ensembles evolved in this fixed potential both with and without low amplitude 
perturbations, including an analysis of what Merritt \& Valluri (1996) have 
termed `chaotic mixing' (cf. Kandrup \& Mahon 1994a, Mahon {\it et al.} 1995, 
Kandrup 1998b). The aim is to understand both (i) the extent to which 
topological bottlenecks like an Arnold (1964) web can impede phase space 
transport in the unperturbed phase space and (ii) how even low amplitude 
irregularities can help orbits to traverse these bottlenecks. Assessing  
these topological effects is crucial for understanding the dynamics of 
potentials that admit a coexistence of both regular and chaotic orbits; and,
as such, an important first step in determining whether it be reasonable to
use them as viable candidates for collisionless equilibria.

Basic questions to be addressed include the following:
\par\noindent${\bullet}$
To what extent does the efficiency of phase space transport depend on the 
steepness of the central cusp? In particular, does steepening the cusp, which 
appears to increase the overall importance of chaos (cf. Merritt \& Fridman 
1996), also make phase space transport more efficient?
\par\noindent${\bullet}$
How is chaotic mixing impacted by low amplitude perturbations, modeled
as friction and noise or periodic driving? In particular, how large do such
perturbations have to be in order to have significant effects within a Hubble
time $t_{H}$? Earlier work on generic complex potentials would suggest that 
even perturbations so weak as to be characterised by a relaxation time 
$t_{R}{\;}{\sim}{\;}10^{6}t_{D}$ or longer can be important dynamically within 
$100t_{D}$ (Habib, Kandrup, \& Mahon 1997).
\par\noindent${\bullet}$
How are bulk, statistical properties altered by the presence of a supermassive 
black hole? Earlier work suggests that a supermassive black hole can increase
the overall abundance of chaotic orbits, and that it may render chaotic orbits
more unstable, i.e., endow them with a larger Lyapunov exponent (cf. Udry \&
Pfenniger 1988, Hasan \& Norman 1990), but it is not clear whether the 
presence of a black hole accelerates or impedes phase space transport. 
\par\noindent
The answers to these questions will then be used to speculate about an even
more fundamental issue, namely:
does it seem reasonable to expect that these, and similar, potentials can be 
used to construct self-consistent (near-)equilibria?

Section 2 recalls some basic results from nonlinear dynamics critical to
a proper understanding of chaotic potentials with a complex phase space, 
neglecting the effects of a cusp and/or a supermassive black hole. Section 3
then describes how, at least for the generalised Dehnen potentials, the 
insertion of a central cusp appears to alter the basic picture. Section 4
focuses on the stability of flows in these cuspy triaxial potentials 
towards low amplitude irregularities. Section 5 then considers how the picture
is complicated by the addition of a supermassive black hole. Section 6 
concludes by summarising the principal results and speculating on their
implications. 

\section{PHASE SPACE TRANSPORT IN CHAOTIC HAMILTONIAN SYSTEMS}

The phase space associated with a Hamiltonian system admitting only regular
or only chaotic orbits tends to be comparatively simple topologically. However,
the coexistence of large measures of both regular and chaotic orbits leads
generically to a complex phase space, the chaotic phase space regions being
laced with a complicated pattern of cantori (cf. Percival 1979) or an Arnold
(1964) web. The former arise for three-degree-of-freedom systems with two
global isolating integrals, e.g., axisymmetric configurations (as well as
two-degree-of-freedom systems with only one isolating integral); the latter for
three-degree-of-freedom systems with only one isolating integral.

It is often -- but not always -- true that, for fixed values of the global 
isolating integrals, the chaotic phase space region is connected in the sense 
that a single chaotic orbit will eventually pass arbitrarily close to every 
point (strictly speaking [cf. Lichtenberg \& Lieberman 1992], this neglects 
tiny chaotic regions nested inside {\it KAM} tori). However, in many 
cases one still finds that, over surprisingly long time scales, literally 
hundreds of dynamical times ($t_{D}$) or longer, the phase space is {\it de 
facto} divided into nearly disjoint regions 
by cantori or Arnold webs, which can serve as efficient bottlenecks. 
It follows that, over time scales of astronomical interest, an ensemble of 
chaotic orbits, all with the same energy, can -- even if there are no other 
isolating integrals -- divide into two or more distinct populations, 
distinguished by (i) what parts of the chaotic phase space they occupy and/or 
(ii) how chaotic  they are (cf. Mahon {\it et al.} 1995).
This phenomenon can be probed and quantified using tools like short time 
Lyapunov exponents (cf. Kandrup \& Mahon 1994a), which were first introduced in
a mathematically precise setting by Grassberger {\it et al.} (1988) or,
alternatively, by probing the extent to which the Fourier spectrum has 
broad band power (Kandrup, Eckstein, \& Bradley 1997). 

Within a given phase space region unobstructed by major cantori or the Arnold 
web, chaotic mixing is usually quite efficient (Kandrup \& Mahon 1994a, Merritt
\& Valluri 1996). In particular, for an initially localised ensemble of orbits 
one observes typically (i) an initial exponential divergence in phase space, 
followed by (ii) an exponential approach towards a near-uniform population of 
the phase space region, a so-called near-invariant distribution. Both these 
effects proceed rapidly on a mixing time scale $t_{M}$ comparable to the 
natural time scale $t_{L}=1/{\chi}$ associated with the largest Lyapunov 
exponent. However, cantori and Arnold webs can dramatically increase the time 
required to approach a true invariant distribution (Kandrup 1998b). 
Separate phase space regions reach an `equilibrium' comparatively quickly, but 
the time scale $t_{eq}$ for the different regions to communicate and reach a 
global equilibrium can be extremely long!

The notion of an {\it invariant distribution} (cf. Lichtenberg \& Lieberman
1992), corresponding to a microcanonical equilibrium, i.e., a uniform 
population of the accessible phase space regions, plays a crucial role in any 
self-consistent equilibrium. When evolved into the future, a generic initial 
phase density $f(0)$ will be transformed into a new $f(t){\;}{\ne}{\;}f(0)$ 
and, as such, cannot serve as a time-independent building block. However, a 
uniform phase density $f_{\rm inv}$ is invariant under an evolution governed 
by Hamilton's equations, i.e., $f_{\rm inv}(t)=f_{\rm inv}(0)$. Such an
$f_{\rm inv}$ thus constitutes a time-independent building block that can be 
used for constructing a self-consistent equilibrium.

This is especially important for triaxial systems which typically admit only 
one global isolating integral (cf. Binney \& Tremaine 1987). Considering
phase space distributions that depend on one global integral and nothing
else seems too restrictive to model centrally condensed
triaxial systems. (For example, any nonrotating equilibrium that depends only
on energy must (cf. Perez and Aly 1996) be spherical.) This means that, if a 
triaxial potential admitting 
only one global integral is to be used to construct an equilibrium model, that
equilibrium must involve `nonstandard' time-independent building blocks, e.g., 
with different weights assigned to regular and chaotic orbits with the same 
energy (cf. Kandrup 1998a). The important point then is that these building
blocks must be truly time-independent if the distribution is to correspond to 
a true equilibrium. 

One could envision near-invariant distributions $f_{\rm niv}$ 
corresponding to nearly constant populations of chaotic phase space regions 
that are separated from the rest of the chaotic phase space by cantori or an 
Arnold web and, consequently, are nearly time-independent. However, as
time elapses orbits will diffuse through the surrounding obstructions and 
$f_{\rm niv}$ will approach the true invariant distribution, $f_{\rm inv}$.

The possibility of distinguishing between `sticky' and `not sticky' chaotic
orbit segments was recognised nearly thirty years ago (cf. Contopoulos 1971).
In particular, it was observed that a chaotic segment can be stuck
near a regular island and behave as if it were regular for very long times, 
hundreds of dynamical times or longer, even though it will eventually become 
unstuck. Such near-regular chaotic orbits can play an important role in 
galactic modeling (cf. Athanassoula {\it et al.} 1983, Wozniak 1993, Kaufmann 
\& Contopoulos 1996). Conventional wisdom suggests that regular orbits
must provide the skeleton for a self-consistent model (cf. Binney 1978). 
However, in certain critical phase space regions (e.g., near corotation) almost
no regular orbits may exist, although sticky orbits are still abundant. Using 
near-regular sticky orbits seems a logical alternative.

In principle this is completely reasonable, but there is an important
potential concern: even very weak perturbations can dramatically accelerate 
phase space transport through cantori or along an Arnold web, 
allowing sticky orbits to become unsticky and vice versa (cf. Lieberman \& 
Lichtenberg 1972, Lichtenberg \& Wolf 1989, Habib, Kandrup, \& Mahon 1997).
One obvious perturbation is discreteness effects which (cf.
Chandrasekhar 1943a), are usually modeled as dynamical friction and white
noise (e.g., in the context of a Fokker-Planck description). For regular 
orbits in a fixed potential, such friction and noise only have significant 
effects on a collisional relaxation time $t_{R}$ which, for galaxies as a 
whole, is usually long compared with $t_{H}$. However, such
perturbations {\it can} have significant effects on the statistical properties 
of chaotic orbit ensembles on much shorter times by accelerating diffusion 
through cantori (Habib, Kandrup, \& Mahon 1997) or along an Arnold web 
(Kandrup, Pogorelov, \& Sideris 2000). 

Another class of perturbations reflects companion objects/nearby 
galaxies and internal pulsations which, in at least some cases, can be modeled 
by a (near-) periodic driving. Not surprisingly, such 
driving is most effective when the driving frequencies are comparable to 
the natural frequencies of the unperturbed orbits (cf. Kandrup, Abernathy, \& 
Bradley 1995). However, driving can be important even if the driving 
frequencies are quite low compared with the natural frequencies (cf. 
Lichtenberg \& Lieberman 1992), the resonant coupling in this case arising via 
subharmonics. Both these  phenomena are known to be important in the physics 
of nonneutral plasmas (cf. Tennyson 1979, Rechester, Rosenbluth, \& White 1981,
Habib \& Ryne 1995).

In a rich cluster, the external environment probably cannot be 
modeled simply as a superposition of a small number of near-periodic forces. 
And similarly, there may be internal irregularities that cannot be approximated
reasonably as nearly periodic. However, one might still anticipate that these 
effects can be modeled as a `random' (i.e., stochastic) process involving a 
superposition of a large number of different frequencies (formally, any signal 
can be decomposed into a sum of periodic Fourier components). For a 
random combination of frequencies combined with random phases, this is 
equivalent mathematically to (in general) coloured noise (cf. van Kampen 1981,
Honerkamp 1994), with a finite autocorrelation time and a band-limited power 
spectrum.

It is significant that evidence for irregular 
shapes, as probed by isophotal twists or $\cos 3\theta$ corrections to 
isophotal ellipses, is especially common in high density environments (cf. 
Zepf \& Whitmore 1993, Mendes de Olivera \& Hickson 1994). This suggests 
that such irregularities could be induced environmentally by collisions and
other close encounters; and, even more fundamentally, that these galaxies have
been displaced from equilibrium by their surrounding environment. At the most
extreme level, these observations could raise the question of whether it be
reasonable to model galaxies as self-consistent equilibria. A more conservative
response is to determine whether such self-consistent equilibria are stable 
towards low amplitude irregularities.

The crucial point in all this is that even if, in the absence of all 
perturbations, a galactic model behaves as a stable or near-stable equilibrium
for times $t>t_{H}$, it may be destabilised within $t<t_{H}$ by comparatively 
weak but `realistic' perturbations associated with internal irregularities, 
systematic pulsations, or the surrounding environment. 

The effects of these perturbations do not turn on abruptly: there is no 
obvious critical amplitude below which they are irrelevant. Instead, one finds 
that the efficacy of the perturbations typically scales roughly 
logarithmically with amplitude. Why do these perturbations have an effect? 
Periodic driving is easily understood as inducing a resonant coupling between 
the driving frequencies and the natural frequencies of the unperturbed orbit. 
The driving is comparatively ineffectual if these frequencies are extremely 
different, although one {\it can} get couplings via sub- and superharmonics. 
Noise-induced phase space transport can also be understood as a resonance 
phenomenon. Coloured noise with a band-limited power spectrum only has an 
appreciable effect if the noise has substantial power at frequencies comparable
to the natural orbital frequencies. When the autocorrelation time $t_{c}$ is 
shorter than, or comparable to, $t_{D}$, so that there is appreciable power at 
frequencies ${\sim}{\;}1/t_{D}$, coloured noise has almost the same effect
as white noise. As $t_{c}$ increases to larger values, so that the power 
spectrum cuts off at lower frequencies, the efficacy of the noise decreases
logarithmically (Pogorelov \& Kandrup 1999, Kandrup, 
Pogorelov, \& Sideris 1999). 

\section{UNPERTURBED DEHNEN POTENTIALS}

When considering orbits in a nonintegrable potential there are at least three 
distinct notions of `how chaotic,' each of which is relevant to galactic
dynamics.
\pn 1) What fraction of the accessible phase space is chaotic and what 
fraction regular? 
Do there exist the requisite orbit families for the skeleton of a 
self-consistent model? Short of actually building a self-consistent 
distribution function there is in general no guarantee that any given
potential can serve as an equilibrium -- hence the importance of techniques
like Schwarzschild's (1979) method (see also Schwarzschild 1993). However, 
most dynamicists would 
probably agree that potentials for which the phase space contains only a 
very small measure of regular orbits are unlikely candidates for equilibria.
\pn 2) How unstable are individual chaotic orbit segments, i.e., how large
are the (short time) Lyapunov exponents? This question is important for 
at least two reasons: The size of these exponents regulates
the rate at which nearby orbits diverge and, hence, the rate of chaotic
mixing. Moreover, segments with especially small short time Lyapunov
exponents may be nearly indistinguishable from regular orbits over times
${\sim}{\;}t_{H}$, so that one might perhaps use them in lieu of regular 
orbits when modeling complex structures. 
\pn
These first two points are more or less obvious. A phase space can be `very
chaotic' in the sense that the Lyapunov exponents are very large but,
nevertheless, `not so chaotic' in the sense that the relative measure of
chaotic orbits is comparatively small. Less trivial, perhaps, is the 
following:
\pn 3) Are most of the chaotic phase space regions at fixed energy connected 
in the sense that a single chaotic orbit can, and will, access the entire 
region? (Even neglecting tiny chaotic regions nested inside {\it KAM} tori,
there is no guarantee that this is so.) And,
assuming that the answer is yes, to what extent can a single orbit pass
unimpeded throughout that region without being obstructed by cantori or an
Arnold web? Such questions related to phase space transport are important in
galactic dynamics because they impact on the overall efficacy of chaotic
mixing and the extent to which it makes sense to use `sticky' chaotic orbits
as near-regular building blocks. 

\subsection{What fraction of the orbits are chaotic?}

To obtain a reasonable estimate of the relative number of regular and chaotic 
orbits, one requires reasonable samplings of initial conditions. These were 
provided by generating orbit libraries similar to, but more complete than, 
those
computed by Merritt \& Fridman (1996) for use in constructing self-consistent 
equilibria. This entailed approximating each model by $20$ constant energy 
shells, corresponding to phase space hypersurfaces containing $1/21$, $2/21$, 
... of the total mass. Attention focused primarily on three shells, namely the
two lowest and the ninth lowest. The two lowest energy shells presumably feel
the cusp most acutely; the ninth shell, corresponding to an intermediate 
energy, should be dominated less completely by the cusp. For each choice of
$\gamma$ and energy $E$, orbits were generated for $1000$ initial conditions,
these corresponding closely to the classes of initial conditions originally 
considered by Schwarzschild (1979) (see also Schwarzschild 1993). $250$ of the 
initial conditions had 
${\bf v}{\;}{\equiv}{\;}0$ and, in the absence of chaos, would correspond 
presumably to box orbits. Another $250$ initial conditions were chosen along 
each of the three principal planes, with the two components of velocity in the
plane vanishing identically but the third component 
nonvanishing. These yielded orbits which, in the absence of chaos, might be 
expected to correspond to long, short, and intermediate  axis tubes.

Each orbit was 
integrated for a total time $t=200t_{D}$ and an approximation to the largest
Lyapunov exponent obtained by tracking simultaneously a small initial 
perturbation that was renormalised periodically in the usual way (cf. 
Lichtenberg \& Lieberman 1992). If the chaotic orbits were not `sticky,' one 
would expect such integrations to yield a relatively clean separation between 
regular and chaotic behaviour, integrations with other potentials indicating 
that, in this case, integrating for $t=100t_{D}$ is usually adequate to 
distinguish between chaos and regularity (cf. Kandrup, Pogorelov, \& 
Sideris 1999). However, for these triaxial Dehnen potentials such is not the 
case. In this case, segments of chaotic orbits, evaluated for times 
$t=200t_{D}$ (or even much longer), exhibit invariably a broad range of short 
time Lyapunov exponents, extending from values of ${\chi}$ no larger 
than what is expected for regular orbits to substantially larger values.
For this reason, there can be some uncertainty in determining whether any 
given orbit is, or is not, chaotic. However, it {\it is} possible to determine
the number of ``strongly chaotic'' orbits with short time Lyapunov exponents 
larger than the near-zero values of ${\chi}$ computed for regular orbits.
For the remainder of this subsection the appellation ``chaotic'' refers to 
such strongly chaotic orbits. The quoted values thus represent lower bounds on
the actual number of chaotic orbits.

For ${\gamma}=0$, there seem to be almost no (${\sim}{\;}12$ out of $1000$) 
chaotic 
orbits in the lowest energy shell. Approximately 10\% of the orbits in the 
second shell are chaotic and the number increases to about 15\% in the ninth 
shell. The near-absence of chaos in the lowest energy shell reflects the fact 
that, because there is no cusp, the very lowest energy orbits should be 
near-integrable boxes evolving in what is essentially an anisotropic harmonic
potential. For ${\gamma}=0.3$, the relative fraction of chaotic orbits is 
quite similar except for the fact that, even in the lowest energy shell, about
5\% of the orbits are chaotic. It is natural to attribute this small increase 
in the fraction of chaotic orbits to the fact that the smooth core has been 
replaced by a cusp, albeit a cusp sufficiently weak that the force acting 
on a star at $r\to 0$ does not diverge. This interpretation is corroborated by 
the fact that, for ${\gamma=0.3}$, chaotic orbits in all three shells
tend invariably to follow orbits that bring them close to the center. For 
example, the minimum distance from the origin for orbits in the lowest energy 
shell varies between $r_{min}=0.003$ and $r_{min}=0.227$, but {\it all} the 
seemingly chaotic orbits with ${\chi}(t=200t_{D})>0.008$ have $r_{min}<0.07$ 
and most have $r_{min}<0.04$. 

The same trend is also observed for the larger values of ${\gamma}$, although
the details are somewhat different. For ${\gamma}=1$, a significant fraction 
of the orbits is chaotic for all three energies but in this case the 
relative number of chaotic orbits {\it decreases} with increasing energy. Here 
approximately 25\% of the lowest energy orbits are chaotic whereas the number 
decreases to ${\sim}{\;}15\%$ for orbits in the ninth shell, roughly the same
as for ${\gamma=0}$ and $0.3$. For ${\gamma}=2$, the relative fraction of 
chaotic orbits decreases from about 30\% in the lowest energy shell to 25\% in 
the ninth shell. This decrease presumably reflects the fact that, for 
${\gamma}=1$ and ${\gamma}=2$, the central cusp plays a dominant role in 
generating chaos. (Recall that the force diverges at $r\to 0$ for 
${\gamma}{\;}{\ge}{\;}1$.) 

To say that much of the chaos is triggered by a `close encounter'
with the central cusp seems an oversimplification. For all three nonzero 
values of ${\gamma}$, there exist significant numbers of orbits with small 
$r_{min}$ that are unquestionably regular. Moreover, even for the cuspless 
model with ${\gamma}=0$, the chaotic orbits tend to have relatively small 
values of $r_{min}$. More reasonable is Merritt's (1996) argument that, at 
least for the lower energy shells, much of the chaos is triggered by an 
appropriate resonance overlap, which is stronger for steeper cusps. 

The relative numbers of orbits with different values of ${\chi}$ can be gauged
from FIGURES 13 and 14, which will be discussed more carefully in Section 5.

\subsection{How unstable are these chaotic orbits?}

To appreciate the significance of this chaotic behaviour, it is important also 
to assess the natural time scale associated with the exponential instability
of these chaotic orbits, both in absolute units and expressed dimensionlessly 
in units of the dynamical time, $t_{D}$. 
The former provides information about the time scale on which the effects of 
chaos could be manifested physically, e.g., in the context of chaotic mixing. 
The latter ties the chaos more securely to the underlying dynamics. An 
estimate of the true Lyapunov exponent, defined in an asymptotic $t\to\infty$ 
limit, was obtained for each value of ${\gamma}$ and $E$ by computing ${\chi}$ 
for each of $8$ chaotic initial conditions with given ${\gamma}$ and $E$, 
integrated for a total time $t=102400t_{D}$, and then constructing the average 
value ${\langle}{\chi}{\rangle}$, as well as dimensional and dimensionless 
time scales $t_{L}=1/{\langle}{\chi}{\rangle}$ and 
$T_{L}=1/({\langle}{\chi}{\rangle}t_{D})$. These integrations were performed
using a variable time-step integrator which typically conserved energy to at
least one part in $10^{5}$ or better. 

The principal conclusion of these computations is that, even though the 
dimensional Lyapunov time $t_{L}$ varies enormously
for different choices of ${\gamma}$ and $E$, the dimensionless $T_{L}$
does not. For the twelve different samples
-- four choices of ${\gamma}$ and three choices of $E$ -- the computed values
of $T_{L}$ varied by less than a factor of $2.5$. In every case the Lyapunov
time on which small perturbations grow exponentially is of order $3-7t_{D}$. 
This is consistent with work in other potentials, in both two and three
dimensions, where, for systems exhibiting global stochasticity, the Lyapunov 
time is typically comparable to, but somewhat longer than, a characteristic 
crossing time (cf. Mahon {\it et al.} 1995, Kandrup, Pogorelov, \& Sideris 
1999). This suggests, however, that chaos should
be {\it much} more important physically at low energies and/or for steep cusps,
where the dynamical time $t_{D}$ is comparatively short. Assuming, e.g., that
the mass $M=5\times 10^{11}M_{\odot}$ and $a=5$ kpc, one finds 
(cf. eq. [13] in Merritt \& Fridman 1996) that, for ${\gamma}=2$, the 
dynamical time for the lowest energy shell is as short as $1.2\times 10^{6}$ 
yr, whereas, for the ninth energy shell for ${\gamma}=1$, 
$t_{D}=8.8\times 10^{7}$ yr.

\begin{figure}
\centering
\centerline{
        \epsfxsize=8cm
        \epsffile{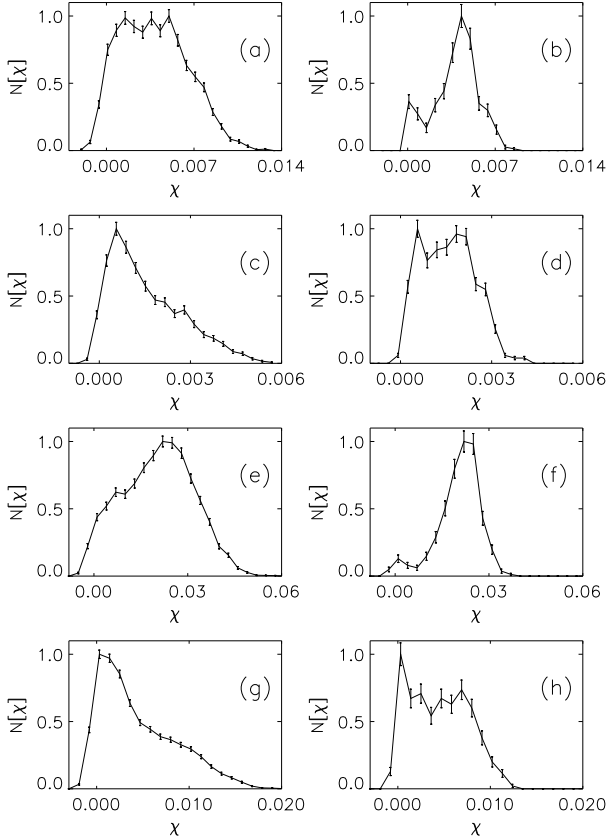}
           }
        \begin{minipage}{10cm}
        \end{minipage}
        \vskip -0.3in\hskip -0.0in
\caption{ 
(a) $N[{\chi}({\Delta}t)]$, the distribution of short time 
Lyapunov exponents, for chaotic orbits in the lowest energy shell with 
${\gamma}=0$ for sampling time ${\Delta}t=100t_{D}$. The error bars reflect
formal $N^{1/2}$ uncertainties in the number of orbits in each bin.
(b) $N[{\chi}({\Delta}t)]$ for the same orbits for ${\Delta}t=800t_{D}$.
(c) $N[{\chi}({\Delta}t)]$ for chaotic orbits in the ninth energy shell with
${\gamma}=0$ and ${\Delta}t=100t_{D}$. 
(d) $N[{\chi}({\Delta}t)]$ for the same orbits for ${\Delta}t=800t_{D}$.
(e) - (h) The same as (a) - (d) for ${\gamma}=0.3$.}
\label{landfig}
\end{figure}

\subsection{Phase space transport}

Visual inspection of the long time trajectories of the chaotic orbits described
above suggests
that (most of) the chaotic portions of the phase space at any given energy are
connected in the sense that a single orbit can and (presumably) eventually 
will pass arbitrarily close to every point in the region. However, visual
inspection of these trajectories also indicates that, for all four values
of ${\gamma}$, but especially the larger values, chaotic orbits can be 
extremely `sticky', even when viewed 
over intervals as long as $10000t_{D}$ or more. Sometimes the orbit segment 
will closely resemble a near-regular box or tube; sometimes it will be wildly
chaotic. These differences are manifested in the behaviour of short time
Lyapunov exponents, which can differ significantly for very long times.
As in other potentials (cf. Kandrup, Eckstein, \& Bradley 1997), one discovers 
that segments that look nearly regular
tend to have relatively small short time ${\chi}$'s, whereas segments that 
are manifestly irregular tend to have larger values of ${\chi}$. The net 
result is that, for a broad range of sampling intervals ${\Delta}t$, the
distribution of short time Lyapunov exponents, $N[{\chi}({\Delta}t)]$, 
generated from the aforementioned long time integrations, is distinctly 
bimodal. 

As is evident from Section 2, this fact is not surprising.
However, there {\it is} at least one important respect in which chaotic 
segments computed in these Dehnen potentials differ from, e.g., orbits in the 
generalised dihedral potential explored by Kandrup, Pogorelov, \& Sideris 
(1999). In that potential, as well as many other generic potentials, 
one finds that, for intervals ${\Delta}t>1000t_{D}$ or so, the distinctions 
between `sticky' and `wildly chaotic' tend to be erased, so that different 
chaotic segments with the same energy have similar statistical properties. In
particular, even if for small ${\Delta}t$ the distribution 
$N[{\chi}({\Delta}t)]$ is bimodal, for ${\Delta}t{\;}{\gg}{\;}1000t_{D}$ the
distribution tends to be singly peaked. For these triaxial Dehnen potentials
this is no longer the case. Even ${\Delta}t=20000t_{D}$ is not long enough for 
these distinctions to be erased. 

The fact that $N[{\chi}({\Delta}t)]$ can be bimodal for comparatively long 
times is illustrated in FIGURES 1 and 2 which, for a variety of different 
values of ${\gamma}$ and $E$, exhibits $N[{\chi}({\Delta}t)]$ for both 
${\Delta}t=100t_{D}$ and ${\Delta}t=800t_{D}$. 

\begin{figure}
\centering
\centerline{
        \epsfxsize=8cm
        \epsffile{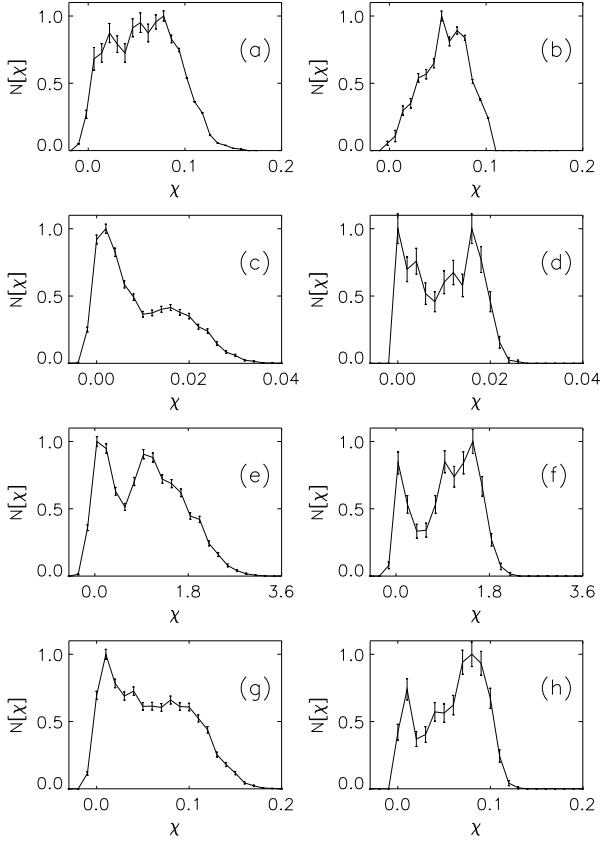}
           }
        \begin{minipage}{10cm}
        \end{minipage}
        \vskip -0.3in\hskip -0.0in
\caption{
(a) $N[{\chi}({\Delta}t)]$, the distribution of short time 
Lyapunov exponents, for chaotic orbits in the lowest energy shell with 
${\gamma}=1$ for sampling time ${\Delta}t=100t_{D}$. 
(b) $N[{\chi}({\Delta}t)]$ for the same orbits for ${\Delta}t=800t_{D}$.
(c) $N[{\chi}({\Delta}t)]$ for chaotic in orbits in the ninth energy shell with
${\gamma}=1$ and ${\Delta}t=100t_{D}$. 
(d) $N[{\chi}({\Delta}t)]$ for the same orbits for ${\Delta}t=800t_{D}$.
(e) - (h) The same as (a) - (d) for ${\gamma}=2$.}
\label{landfig}
\end{figure}

\begin{figure}
\centering
\centerline{
        \epsfxsize=8cm
        \epsffile{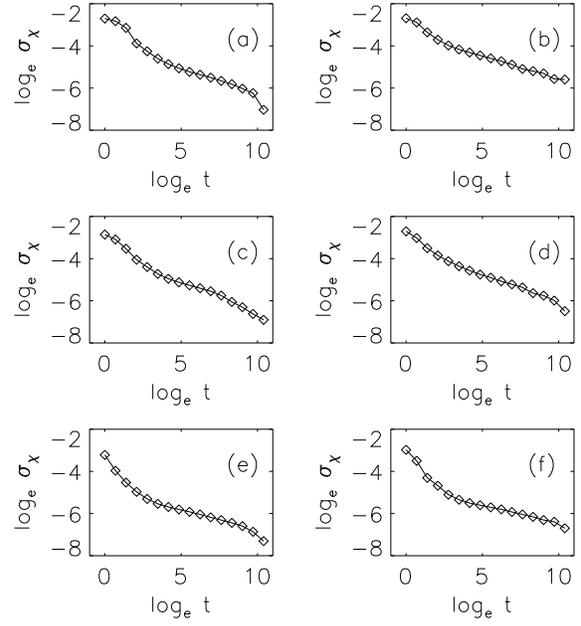}
           }
        \begin{minipage}{10cm}
        \end{minipage}
        \vskip -1.3in\hskip -0.0in
\caption{
(a) The dispersion ${\sigma}_{\chi}$ for chaotic orbits in 
the lowest energy shell with ${\gamma}=0$, computed as a function of sampling 
time ${\Delta}t$. (b) ${\sigma}_{\chi}$ for chaotic orbits in the lowest 
energy shell with ${\gamma}=0.3$. 
(c) The second energy shell with ${\gamma}=0.$
(d) The second energy shell with ${\gamma}=0.3.$
(e) The ninth energy shell with ${\gamma}=0.$
(f) The ninth energy shell with ${\gamma}=0.3.$}
\label{landfig}
\end{figure}
For significantly larger values of ${\Delta}t$, it becomes prohibitively
expensive computationally to compute enough orbit segments to generate a
meaningful distribution of short time ${\chi}$'s. However, it {\it is} still
possible to extract useful information about the distribution by determining
how the dispersion ${\sigma}_{\chi}$ varies as a function of ${\Delta}t$. 
As discussed more carefully in Kandrup, Pogorelov, \& Sideris (1999) (see FIG. 
3 in that paper, as well as Kandrup \& Mahon 1994b), an application of the 
Central Limits Theorem (cf. Chandrasekhar 1943b, van Kampen 1981) suggests 
that if, over the time scales of interest, chaotic orbit segments constitute a 
single population with the overall instability of the orbit at any two instants
essentially uncorrelated, ${\sigma}_{\chi}$ should decrease as 
${\Delta}t^{-1/2}$, whereas the existence of multiple populations implies a 
dispersion that decreases more slowly. The principal conclusion here is that, 
for chaotic orbits in these generalised Dehnen potentials, ${\sigma}_{\chi}$ 
typically decreases {\it very} slowly, {\it much} 
more slowly than ${\Delta}t^{-1/2}$. For ${\Delta}t<2^{5}t_{D}=32t_{D}$ or 
so, ${\sigma}_{\chi}$ typically decreases in a fashion roughly consistent with 
a ${\Delta}t^{-1/2}$ dependence, but for $2^{5}t_{D}<t<2^{17}t_{D}$, i.e.,
$32t_{D}<t<65536t_{D}$, ${\sigma}_{\chi}$ decreases much slower. One
thus anticipates that, consistent with the fact that, even for periods 
${\gg}{\;}10000t_{D}$, different segments exhibit significant variability in
their visual appearance, there exist distinct populations that persist for
$t>10000t_{D}$. FIGURES 3 and 4 exhibit ${\sigma}_{\chi}({\Delta}t)$ for a 
variety of different values of ${\gamma}$ and $E$.

\begin{figure}
\centering
\centerline{
        \epsfxsize=8cm
        \epsffile{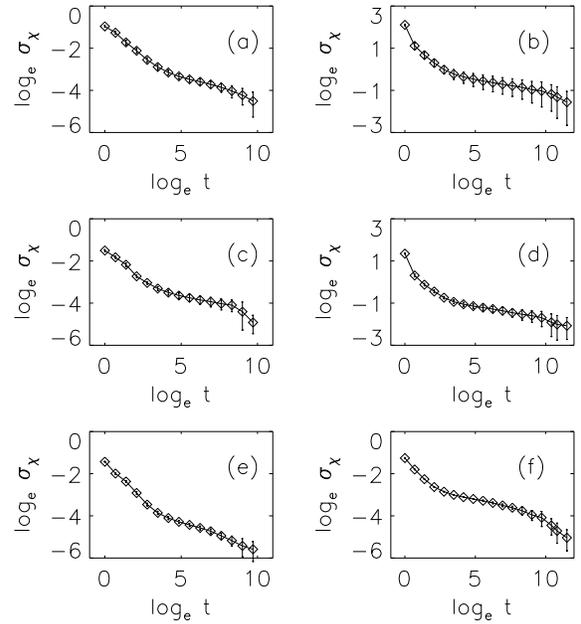}
           }
        \begin{minipage}{10cm}
        \end{minipage}
        \vskip -1.3in\hskip -0.0in
\caption{
(a) The dispersion ${\sigma}_{\chi}$ for chaotic orbits in 
the lowest energy shell with ${\gamma}=1$, computed as a function of sampling 
time ${\Delta}t$. (b) ${\sigma}_{\chi}$ for chaotic orbits in the lowest 
energy shell with ${\gamma}=2$. 
(c) The second energy shell with ${\gamma}=1.$
(d) The second energy shell with ${\gamma}=2.$
(e) The ninth energy shell with ${\gamma}=1.$
(f) The ninth energy shell with ${\gamma}=2.$}
\label{landfig}
\end{figure}

\begin{figure}
\centering
\centerline{
        \epsfxsize=8cm
        \epsffile{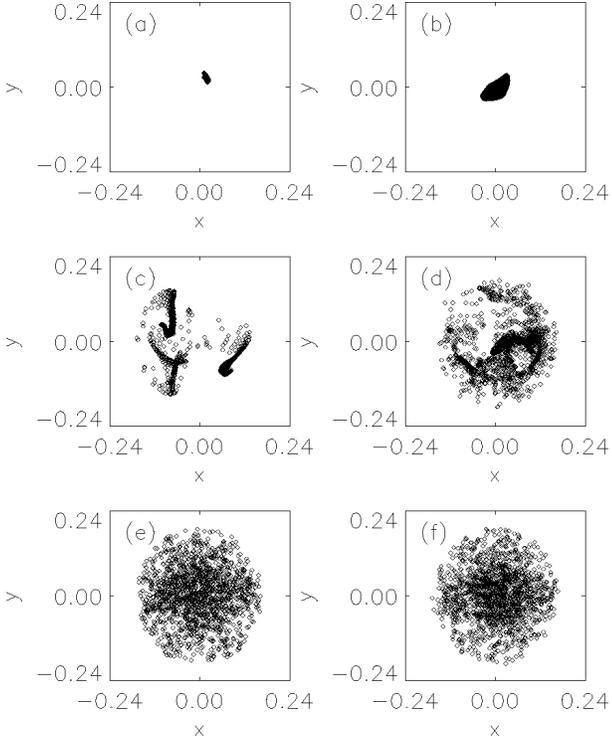}
           }
        \begin{minipage}{10cm}
        \end{minipage}
        \vskip -0.3in\hskip -0.0in
\caption{
$x-y$ coordinates for a $1600$-orbit ensemble evolved with 
${\gamma}=1$ with the lowest energy recorded at six different times. (a) 
$t=t_{D}$. (b) $t=4t_{D}$. (c) $t=8t_{D}$. (d) $t=16t_{D}$. (e) $t=40t_{D}$. 
(f) $t=100t_{D}$.}
\label{landfig}
\end{figure}

\subsection{Chaotic mixing}

These results have obvious implications for chaotic mixing. 
Earlier work (cf. Kandrup \& Mahon 1994a, Merritt \& Valluri 1998, Kandrup 
1998b) would suggest (i) that, as probed by quantities like the phase space
dispersions ${\sigma}_{x}$ and ${\sigma}_{px}$, initially localised ensembles
of chaotic orbits will diverge exponentially, and (ii) that, as probed by 
various lower-order moments and/or coarse-grained reduced distribution 
functions, they will exhibit a roughly exponential evolution towards a 
near-equilibrium (a near-invariant distribution), both effects proceedings 
on a mixing time scale $t_{M}$ that correlates with the Lyapunov time scale 
$t_{L}=1/{\chi}$ associated with the largest Lyapunov exponent. 
However, to the extent that the typical values of short time Lyapunov 
exponents can differ substantially for different ensembles, one might 
anticipate that these ensembles could approach a near-equilibrium at very 
different rates; and, to the extent that orbits can be `stuck' in a given
portion of the chaotic phase space for a very long time, one would not expect
that the near-equilibrium should be independent of the choice of chaotic 
ensemble. Rather, one would anticipate a two- (or more) stage evolution,
whereby a rapid approach towards a uniform population of the easily accessible 
regions is followed by a slower evolution as orbits diffuse along the Arnold 
web to
probe the entire accessible phase space. The approach towards a near-invariant
distribution typically proceeds on a comparatively short time scale $t_{M}
{\;}{\sim}{\;}1/{\chi}$. The final approach towards a true invariant
distribution often proceeds on a time scale $t_{eq}{\;}{\gg}{\;}1/{\chi}$.

These ideas were tested for the triaxial Dehnen potentials by performing 
simulations which tracked the evolution of ensembles of $1600$ initial 
conditions localised within phase space cells of typical size $0.01$ or less,
and these expectations were all confirmed.
Numerical evolution of localised ensembles of initial conditions indicates
that quantities like the dispersion ${\sigma}_{x}$ in the coordinate $x$ 
do indeed diverge exponentially at a rate that correlates with an average 
short time Lyapunov exponent for the ensemble. Moreover, coarse-grained 
reduced distribution functions constructed from these orbits appear to 
converge exponentially towards a nearly constant $f_{\rm inv}$.
However, the rate at which the dispersions grow and the rate of convergence 
towards $f_{\rm inv}$ can both depend on the choice of ensemble as well as the 
direction(s) in phase space that are probed. For example, for the lowest 
energy shell with ${\gamma}=1$, there is often a propensity for ensembles 
starting with small $z$ to diverge comparatively slowly in the $z$-direction,
so that ${\sigma}_{z}$ approaches a near-constant value more slowly than 
${\sigma}_{x}$ and ${\sigma}_{y}$, and the coarse-grained distribution function
converges towards $f_{\rm inv}$ more slowly in the $z$- and $v_{z}$-directions
than in any other direction. This is illustrated in FIGURES 5 and 6, which 
exhibit, respectively, the $x-y$ and $y-z$ coordinates for the same $1600$ 
orbit ensembles at times $t=t_{D}$, $4t_{D}$, $8t_{D}$, $16t_{D}$, $40t_{D}$,
and $100t_{D}$. Alternatively, the spatial dispersions ${\sigma}_{x}$ and
${\sigma}_{z}$ are exhibited in the top two panels of FIGURE 8, which is 
discussed more carefully in the following Section.

\begin{figure}
\centering
\centerline{
        \epsfxsize=8cm
        \epsffile{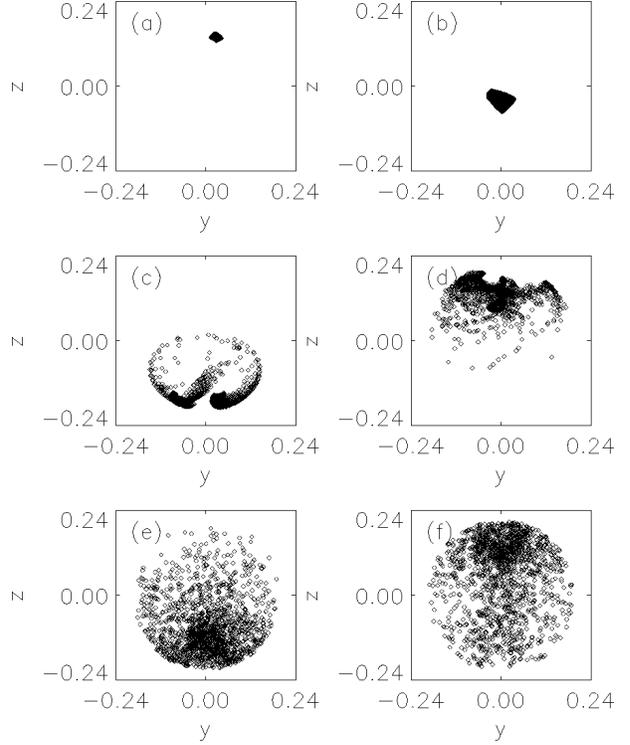}
           }
        \begin{minipage}{10cm}
        \end{minipage}
        \vskip -0.3in\hskip -0.0in
\caption{
$y-z$ coordinates for the same $1600$-orbit ensemble 
recorded at six different times. (a) $t=t_{D}$. (b) $t=4t_{D}$. 
(c) $t=8t_{D}$. (d) $t=16t_{D}$. (e) $t=40t_{D}$. (f) $t=100t_{D}$.}
\label{landfig}
\end{figure}

That the near-invariant distributions generated from different chaotic 
ensembles with essentially the same energy need not be statistically identical 
is illustrated by the fact that the moments associated with different ensembles
need not evolve towards the same near-constant values. An example thereof
is exhibited in the top six panels of FIGURE 7, which compares the 
time-dependent dispersions ${\sigma}_{x}$, ${\sigma}_{y}$, and ${\sigma}_{z}$ 
for two different ensembles of 6400 orbits, each sampling the lowest energy 
shell with ${\gamma}=1$. Both ensembles were generated from initial conditions 
with ${\bf v}=0$, i.e., orbits which, in the absence of a cusp, might be 
expected to correspond to regular boxes. Each panel is indicative of a 
moment converging towards a near-constant value, but it is evident that 
these values are distinctly different for the two different ensembles. 
One might perhaps worry that these differences could reflect the fact that some
of the orbits in these ensembles are actually regular. That this is not the 
case is clear from the last two panels in FIGURE 7, which exhibit distributions
of short time Lyapunov exponents, $N[{\chi}(t)]$, for each of the $6400$-orbit
ensembles, computed at $t=200t_{D}$. In each case, a broad range of values
are assumed, but the distributions are unimodal and the lowest ${\chi}$ is 
significantly displaced from ${\chi}=0$.

\begin{figure}
\centering
\centerline{
        \epsfxsize=8cm
        \epsffile{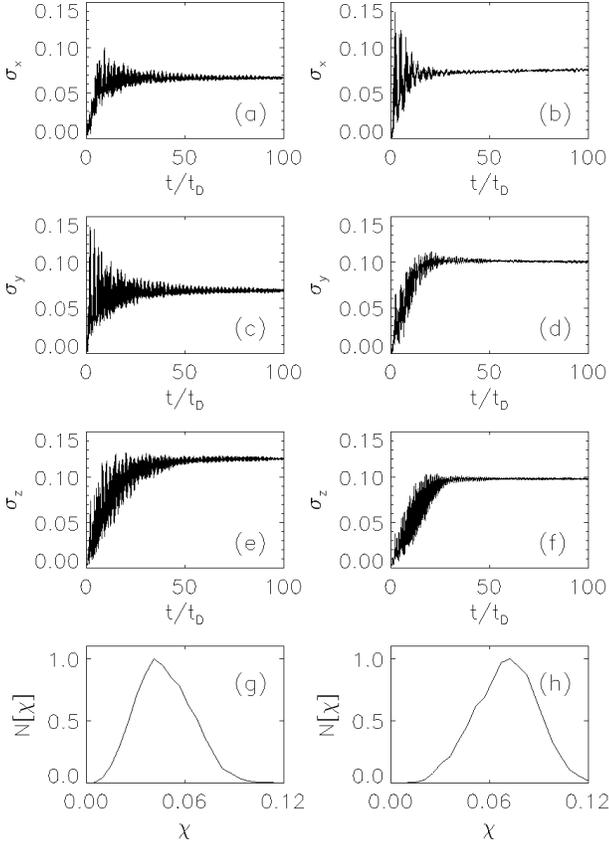}
           }
        \begin{minipage}{10cm}
        \end{minipage}
        \vskip -0.3in\hskip -0.0in
\caption{
(a) The dispersion ${\sigma}_{x}(t)$ generated for one $6400$ orbit chaotic
ensemble sampling the lowest energy shell with ${\gamma}=1$. (b) 
${\sigma}_{x}$ for another chaotic ensemble with the same $E$ and ${\gamma}$.
(c) and (d) ${\sigma}_{y}$ for the same ensembles. (e) and (f) ${\sigma}_{z}$
for the same ensembles. (e) A distribution of largest short time Lyapunov 
exponents, $N[{\chi}(t)]$, for $t=200t_{D}$, generated for the first ensemble.
(f) The same for the second ensemble.}
\label{landfig}
\end{figure}

\section{THE EFFECTS OF INTERNAL AND EXTERNAL IRREGULARITIES}

The objective here is to determine how the statistical properties of chaotic 
orbits evolved in these generalised Dehnen potentials change if the evolution
equations are modified to allow for low amplitude irregularities which could 
mimic the effects of internal substructures and/or an external environment.
This entailed repeating the simulations of chaotic mixing for unperturbed 
systems described in the preceding Section but now allowing for low amplitude
perturbations. Three specific effects were considered:
\par\noindent${\bullet}$
{\it Periodic driving}, intended to mimic the effects of one or two 
companion objects and/or internal pulsations.
This involved allowing for an evolution equation of the form 
\begin{equation}
{d^{2}x^{a}\over dt^{2}}=-{{\partial}V({\bf r})\over {\partial}x^{a}}
+A^{a}\sin ({\omega}_{a}t+{\varphi}_{a}), \qquad (a=x,y,z),
\end{equation}
incorporating simple pulsations in three orthogonal directions with different 
amplitudes, frequencies, and phases. 
\par\noindent${\bullet}$
{\it Friction and additive white noise}, intended to mimic 
discreteness effects, i.e., gravitational Rutherford scattering.
This involved solving a Langevin equation (cf. Chandrasekhar 1943b, 
van Kampen  1981) 
\begin{equation}
{d^{2}x^{a}\over dt^{2}}=-{{\partial}V({\bf r})\over {\partial}x^{a}}
-{\eta}v^{a}+F^{a}, \qquad (a=x,y,z), 
\end{equation}
with ${\eta}$ a constant coefficient of dynamical friction and ${\bf F}$ 
a `stochastic force.' Assuming in the usual fashion that ${\bf F}$ 
corresponds to homogeneous Gaussian noise, its statistical properties
are characterised completely by its first two moments, which take the form
\begin{displaymath}
{\langle}F_{a}(t){\rangle}=0 \qquad {\rm and}
\end{displaymath}
\begin{equation}
{\langle}F_{a}(t_{1})F_{b}(t_{2}){\rangle}=
{\delta}_{ab}\;K(t_{1}-t_{2}),
\hskip .2in (a,b=x,y,z).
\end{equation}
The assumption that the noise be white implies that the autocorrelation 
function $K$ is proportional to a Dirac delta, so that
\begin{equation}
K({\tau})=2{\eta}{\Theta}{\delta}_{D}({\tau}) .
\end{equation}
The normalisation here ensures that the friction and noise are related
by a Fluctuation-Dissipation Theorem at a temperature ${\Theta}$.
\par\noindent${\bullet}$
{\it Coloured noise}, intended primarily to mimic the effects of a stochastic 
external environment. 
This entailed allowing for forces that are random but, nevertheless,
of finite duration, i.e., characterised by a finite autocorrelation time.
The specific example considered here corresponds to the so-called 
Ornstein-Uhlenbeck process (cf. van Kampen 1981), for which $K$ decreases 
exponentially in time, i.e.,
\begin{equation}
K({\tau})={\alpha}{\eta}{\Theta}\,\exp(-{\alpha}|{\tau}|),
\end{equation}
where the autocorrelation time $t_{c}=1/{\alpha}$ sets the time scale
on which ${\bf F}$ changes appreciably. 
The normalisation here ensures that the diffusion constant
$D$ that would enter into a Fokker-Planck description,
\begin{equation}
D=\int_{-\infty}^{\infty}\,d{\tau}K({\tau})=2{\Theta}{\eta},
\end{equation}
is independent of ${\alpha}$.

White noise was implemented computationally using an algorithm described in 
Griner, Strittmatter, \& Honerkamp (1988). Coloured noise was implemented 
using an algorithm similar to that described in Pogorelov \& Kandrup (1999).

This modeling of perturbations is clearly simplistic but, nevertheless, not
completely unreasonable. Earlier work on periodic driving as a source of 
phase space transport suggests strongly that the exact way in which the system
is pulsed matters less than the amplitude and driving frequency. For example,
Kandrup, Abernathy, and Bradley (1995) found that, for variants of an 
anisotropic Plummer potential, similar results obtained when pulsing the core
radius and the anisotropy parameter. 

Although the detailed forms of the friction and noise can be very important
for phenomena like evaporation from a cluster (cf. Chandrasekhar 1943a) or
barrier penetration problems (cf. Lindenberg \& Seshadri 1981), they appear
to be comparatively unimportant in determining the rate of phase space 
transport. Work on simpler two- and three-dimensional systems (Pogorelov \&
Kandrup 1999, Kandrup, Pogorelov, \& Sideris 1999) shows that making ${\eta}$
a simple but nontrivial function of ${\bf v}$ and/or turning off the friction
term altogether has only minimal effects. Moreover, the detailed form of the
coloured noise seems largely immaterial. What really matter are the 
amplitude and the autocorrelation time $t_{c}$, which determines the 
frequencies at which the
perturbation has substantial power. When $t_{c}{\;}{\ll}{\;}t_{D}$, colour 
matters very little and the effects are similar to white
noise. However, when $t_{c}{\;}{\gg}{\;}t_{D}$, so that there is little power
at frequencies ${\sim}{\;}1/t_{D}$, the efficacy of the perturbation 
decreases logarithmically.

\subsection{Friction and white noise}

Consider first discreteness effects, modeled as friction and white noise.
Here the most obvious conclusion is that even weak perturbations can 
significantly accelerate phase space transport by allowing orbits that are
comparatively `sticky' in one or more phase space directions to become
`unstuck.' This is illustrated in FIGURE 8, which exhibits the configuration
space dispersions ${\sigma}_{x}$ and ${\sigma}_{z}$ for the ensemble of initial
conditions used to generate FIGURES 5 and 6, contrasting unperturbed motion
with orbits perturbed by friction and additive white noise with constant
${\Theta}=-E$ and three different values of ${\eta}$, namely ${\eta}=0.4\times 
10^{-7}$, $0.4\times 10^{-6}$, and $0.4\times 10^{-5}$. Given that, for this 
choice of ${\gamma}$ and energy, $t_{D}{\;}{\approx}{\;}2.444$, these values 
correspond respectively to relaxation times $t_{R}=1/{\eta}{\;}{\approx}
{\;}10^{7}t_{D}$, $10^{6}t_{D}$ and $10^{5}t_{D}$. 

\begin{figure}
\centering
\centerline{
        \epsfxsize=8cm
        \epsffile{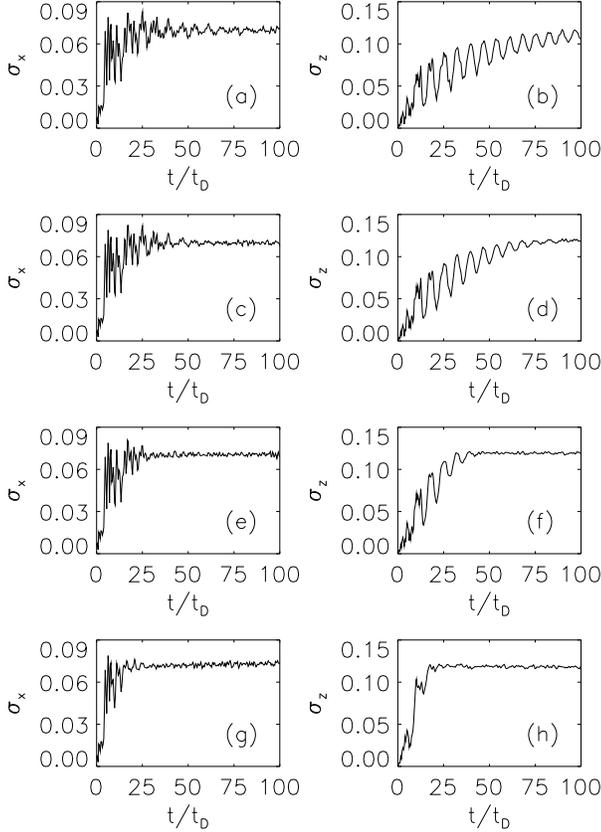}
           }
        \begin{minipage}{10cm}
        \end{minipage}
        \vskip -0.3in\hskip -0.0in
\caption{
(a) The dispersion ${\sigma}_{x}$ computed for an initially
localised ensemble of $1600$ orbits in the lowest energy shell with 
${\gamma}=1$ integrated in the absence of any perturbations. (b) The
corresponding ${\sigma}_{z}$. 
(c) ${\sigma}_{x}$ for the same ensemble, now allowing for friction and white
noise with temperature ${\Theta}=-E$ and amplitude ${\eta}=10^{-7}$.
(d) ${\sigma}_{z}$ for the same friction and noise.
(e) ${\sigma}_{x}$ for the same ensemble, with ${\Theta}=-E$ and 
${\eta}=10^{-6}$.
(f) ${\sigma}_{z}$ for the same friction and noise.
(g) ${\sigma}_{x}$ for the same ensemble, with ${\Theta}=-E$ and 
${\eta}=10^{-5}$.
(h) ${\sigma}_{z}$ for the same friction and noise.}
\label{landfig}
\end{figure}

It is apparent that friction and noise with ${\eta}=0.4\times 10^{-7}$ has a 
relatively small effect, that ${\eta}=0.4\times 10^{-6}$ has a substantially 
larger effect, and that perturbations with ${\eta}=0.4\times 10^{-5}$ suffice
to make ${\sigma}_{z}$ evolve as if the ensemble were hardly sticky at all. 
This behaviour is also manifested by the visual appearance of the orbit 
ensemble, as illustrated in FIGURE 9, which, for ${\eta}=10^{-6}$, exhibits 
the $y$ and $z$ coordinates of the particles at the same times as for FIGURE 6.

\begin{figure}
\centering
\centerline{
        \epsfxsize=8cm
        \epsffile{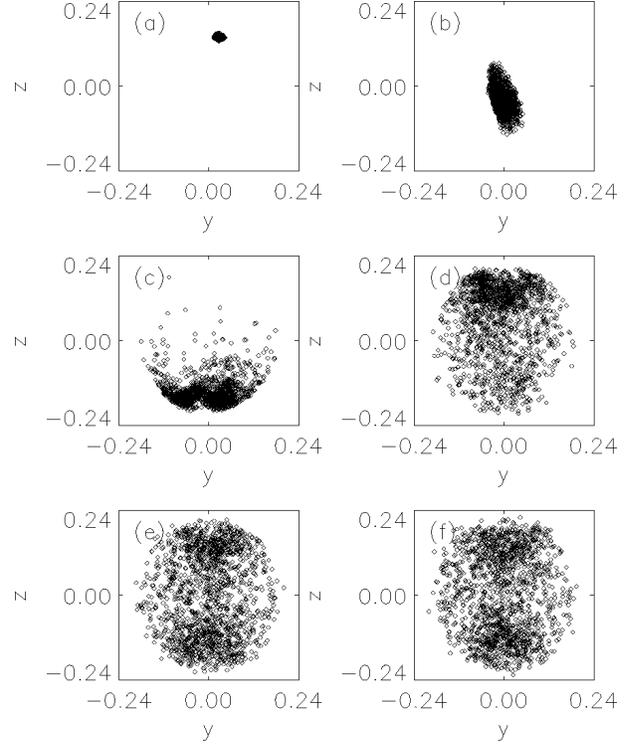}
           }
        \begin{minipage}{10cm}
        \end{minipage}
        \vskip -0.3in\hskip -0.0in
\caption{
$y-z$ coordinates for the $1600$-orbit ensemble of FIGS.
5 and 6, now evolved with friction and white noise with ${\Theta}=-E$ and 
${\eta}=10^{-6}$. (a) 
$t=t_{D}$. (b) $t=4t_{D}$. (c) $t=8t_{D}$. (d) $t=16t_{D}$. (e) $t=40t_{D}$. 
(f) $t=100t_{D}$.}
\label{landfig}
\end{figure}

Even in the absence of stickiness, discreteness effects can be important by
serving to `fuzz out' high frequency structure in quantities like the phase
space dispersions and other lower order moments, as well as various reduced 
distribution functions, thus rendering more efficient the approach towards a 
near-equilibrium. In particular, such noise can serve to damp the oscillations
associated oftentimes with the evolution towards equilibrium, as noted, e.g.,
by Merritt \& Valluri (1996) and Kandrup (1998b). This effect is somewhat 
apparent in FIGURES 8, which exhibit ${\sigma}_{x}$ and ${\sigma}_{z}$ for a 
period $t=100t_{D}$, but is even more striking in FIGURES 10, which exhibit 
${\sigma}_{y}$ for the shorter interval $0<t<50t_{D}$. 

\begin{figure}
\centering
\centerline{
        \epsfxsize=8cm
        \epsffile{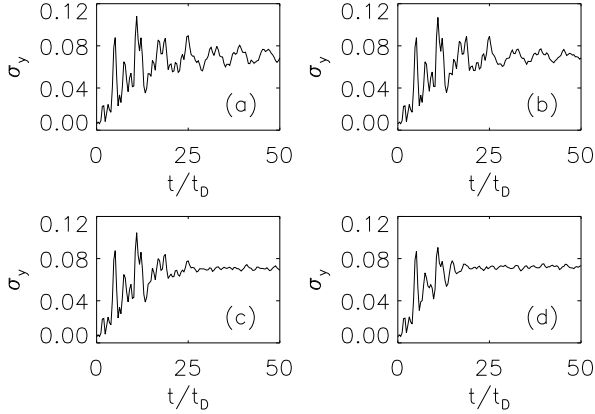}
           }
        \begin{minipage}{10cm}
        \end{minipage}
        \vskip -2.5in\hskip -0.0in
\caption{
(a) The dispersion ${\sigma}_{y}$ computed for an initially
localised ensemble of $1600$ orbits in the lowest energy shell with 
${\gamma}=1$ integrated in the absence of any perturbations. (b) 
${\sigma}_{y}$ for the same ensemble, now allowing for friction and white
noise with temperature ${\Theta}=-E$ and amplitude ${\eta}=10^{-7}$.
(c) ${\sigma}_{y}$ for the same ensemble, with ${\Theta}=-E$ and 
${\eta}=10^{-6}$.
(d) ${\sigma}_{y}$ for the same ensemble, with ${\Theta}=-E$ and 
${\eta}=10^{-5}$}.
\label{landfig}
\end{figure}

How large must ${\eta}$ be to have a significant effect? In general, 
perturbations corresponding to $t_{R}{\;}{\sim}{\;}10^{7}t_{D}$ tend to have 
(at least) a noticeable effect within ${\sim}{\;}20-30t_{D}$, and perturbations
with $t_{R}{\;}{\sim}{\;}10^{6}t_{D}$ can have an appreciable effect on the
overall approach towards a near-equilibrium on a comparable time scale. 
Perturbations with $t_{R}$ as small as $10^{5}t_{D}$ can have drastic effects
within $10t_{D}$ or so. 

Finally, consistent with Habib, Kandrup, \& Mahon
(1997), it was found that, for chaotic ensembles, the root mean squared change
in energy scales as
\begin{equation}
{\delta}E_{rms}{\;}{\approx}{\;}(|E|{\Theta}{\eta}t)^{1/2},
\end{equation}
with $E$ the original energy, which implies that noise is important already 
for ${\delta}E_{rms}{\;}{\sim}{\;}10^{-3}|E|$. This reinforces the 
interpretation
(cf. Pogorelov \& Kandrup 1999, Kandrup, Pogorelov, and Sideris 1999) that 
this phase space transport is {\it not} driven by changes in energy,
which only become large on a relaxation time $t_{R}$, but
instead reflects diffusion along Arnold webs on (nearly) constant energy 
hypersurfaces.

The basic conclusion derived from these investigations is that discreteness 
effects can
be important within a period as short as $50-100t_{D}$ provided only that the
relaxation time is no larger than $10^{6}-10^{7}t_{D}$ or so. This suggests 
strongly that, especially near the centers of cuspy galaxies, discreteness 
effects should be important on time scales considerably shorter than $t_{H}$,
the age of the Universe.

\subsection{Periodic driving}

Overall, periodic driving has the same qualitative effects on orbit ensembles
as does white noise. Periodic driving can enable ``sticky'' orbits to
become unstuck, thus facilitating phase space transport; and, even in the
absence of topological obstructions, driving can help fuzz out high frequency,
short wavelength structures. Not surprisingly, for fixed amplitude $A$ 
periodic driving has the largest effect for driving frequencies
${\omega}_{i}$ with $1/{\omega}_{i}{\;}{\sim}{\;}t_{D}$, 
so that there is a substantial resonant coupling between the
driving frequencies and the natural frequencies of the unperturbed orbits. 

One important difference between noise and periodic driving is that, whereas
noise represents an incoherent process, involving a random superposition of
frequencies, periodic driving with fixed frequencies is a coherent process.
That periodic driving is coherent would suggest that its coupling to orbits
should scale linearly in amplitude, i.e., ${\propto}{\;}A$, a fact that has
been confirmed numerically (cf. Kandrup, Abernathy, \& Bradley 1995). 
Because noise is incoherent, its coupling scales instead as ${\eta}^{1/2}$
(cf. Habib, Kandrup, \& Mahon 1997). For 
$1/{\omega}_{i}{\;}{\sim}{\;}t_{D}$ it is thus natural to expect that, if 
noise requires an amplitude ${\eta}{\;}{\sim}{\;}10^{-p}$ to induce a 
substantial effect within a given time interval, periodic driving will require 
an amplitude $A{\;}{\sim}{\;}10^{-p/2}$ to have a comparable effect. 

Overall, this scaling tends to hold at least roughly true. However, one 
discovers (i) that
the minimum amplitude required to achieve a given effect can exhibit a 
comparatively sensitive dependence on the choice of frequencies, and (ii) that
the ``typical'' effect of periodic driving is slightly weaker than is predicted
by this scaling. These facts are both easily understood. The fact that noise
involves a superposition of many different frequencies implies that the details
tend to ``wash out,'' and that one is more likely to have power at one of the
``more effective'' frequencies.

\subsection{Friction and coloured noise}
As for other two- and three-dimensional potentials, coloured noise sampling 
the Ornstein-Uhlenbeck process has virtually the same effect as does white
noise with the same amplitude, provided only that the autocorrelation time 
$t_{c}$ is short compared with the dynamical time $t_{D}$. However, when
$t_{c}$ becomes comparable to $t_{D}$ the effects of the noise begin to 
decrease significantly; and, for $t_{c}{\;}{\gg}{\;}t_{D}$ the effects of the
noise are substantially reduced. 

Examples of this behaviour, which corroborate
the interpretation of noise-induced phase space transport as a resonance
phenomenon, are presented in FIGURE 11, which exhibits ${\sigma}_{z}$ 
generated from the same ensemble of initial conditions used to generate 
FIGURE 8, now evolved with coloured noise. 
The first three panels represent friction and noise with ${\Theta}=
E$ and ${\eta}=0.4\times 10^{-5}$ for three choices of autocorrelation time, 
namely $t_{c}{\;}{\approx}{\;}0.4t_{D}$, $1.2t_{D}$, and $4.0t_{D}$. The 
fourth and fifth panels were generated for ${\Theta}=-E$ and $t_{c}{\;}
{\approx}{\;}4.0t_{D}$ but larger values of ${\eta}=0.4\times 10^{-4}$ 
and $0.4\times 10^{-3}$. The final panel is generated for ${\Theta}=-E$, 
${\eta}=1.2\times 10^{-3}$, and $t_{c}{\;}{\approx}{\;}12t_{D}$. It is clear
that for these values, which are not inappropriate for modeling the effects of
an external environment on a galaxy embedded in a rich cluster, perturbations
can have a dramatic effect.
\begin{figure}
\centering
\centerline{
        \epsfxsize=8cm
        \epsffile{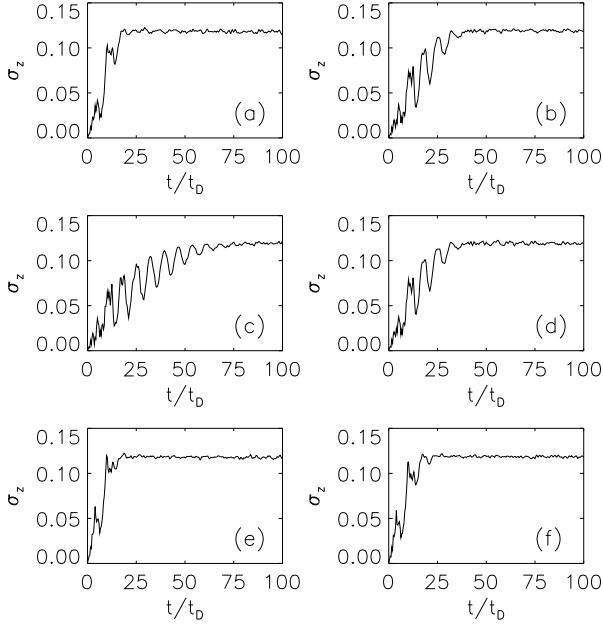}
           }
        \begin{minipage}{10cm}
        \end{minipage}
        \vskip -1.0in\hskip -0.0in
\caption{
(a) The dispersion ${\sigma}_{z}$ computed for the same initial conditions
as FIG. 8, now allowing for coloured noise with ${\Theta}=-E$, 
${\eta}=0.4\times 10^{-5}$, and 
$t_{c}=1/{\alpha}=1.0{\;}{\approx}{\;}0.4t_{D}$.
(b) The same for ${\eta}=0.4\times 10^{-5}$, and 
$t_{c}=3.0{\;}{\approx}{\;}1.2t_{D}.$
(c) The same for ${\eta}=0.4\times 10^{-5}$, and 
$t_{c}=10.0{\;}{\approx}{\;}4t_{D}.$
(d) The same for ${\eta}=0.4\times 10^{-4}$, and 
$t_{c}=10.0{\;}{\approx}{\;}4t_{D}.$
(e) The same for ${\eta}=0.4\times 10^{-3}$, and 
$t_{c}=10.0{\;}{\approx}{\;}4t_{D}.$
(f) The same for ${\eta}=1.2\times 10^{-3}$, and 
$t_{c}=30.0{\;}{\approx}{\;}12t_{D}.$}
\label{landfig}
\end{figure}

\subsection{The approach towards an invariant distribution}
Even though low amplitude perturbations can accelerate the approach towards
a near-invariant distribution, they need not suffice to make different
ensembles evolve towards the same distribution within a time $t{\;}{\sim}{\;}
100t_{D}$ or so. For example, even though friction and noise corresponding
to a relaxation time  as long as $t_{R}{\;}{\sim}{\;}10^{6}t_{D}$ or more can
have an appreciable effect on the efficacy with which the ensemble approaches
a near-invariant distribution, perturbations corresponding to a $t_{R}$ 
as short as $10^{4}t_{D}$ may not suffice to yield a convergence of 
moments for two different chaotic ensembles with the same ${\gamma}$ and $E$. 
This is,
e.g., illustrated in FIGURE 12, which exhibits ${\sigma}_{z}$ for the same
two sets of initial conditions used to generate FIGURE 7, but now allowing for 
friction and white noise with $t_{R}=10^{6}t_{D}$, $10^{5}t_{D}$, and 
$10^{4}t_{D}$. The $t=200t_{D}$ dispersions for the two ensembles are 
appreciably different even for $t_{R}=10^{4}t_{D}$, where the friction and
noise have caused the energy of a typical orbit to change by more than 4\%.
Both ensembles began with energies $E{\;}{\approx}{\;}-0.996$. When evolved
allowing for perturbations corresponding to $t_{R}=10^{4}t_{D}$, the two final 
energy dispersions were ${\sigma}_{E}=0.045$ and ${\sigma}_{E}=0.046$.

\begin{figure}
\centering
\centerline{
        \epsfxsize=8cm
        \epsffile{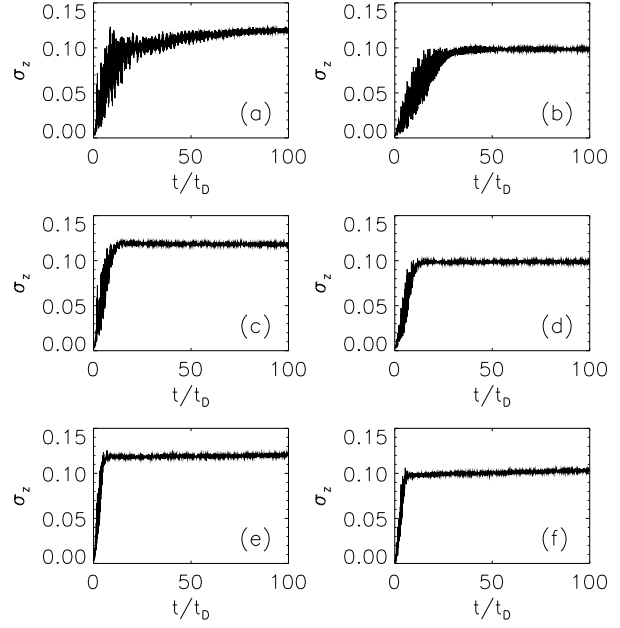}
           }
        \begin{minipage}{10cm}
        \end{minipage}
        \vskip -1.0in\hskip -0.0in
\caption{
(a) The dispersion ${\sigma}_{z}$ computed for one initially
localised ensemble of chaotic orbits in the lowest energy shell with 
${\gamma}=1$, allowing for friction and white noise corresponding
to $t_{R}=10^{6}t_{D}$ (b) ${\sigma}_{z}$ for another chaotic ensemble
with the same ${\gamma}$, $E$ and $t_{R}$. (c) and (d) ${\sigma}_{z}$ for the
same two ensembles of initial conditions, now integrated with 
$t_{R}=10^{5}t_{D}$.(e) and (f) ${\sigma}_{z}$ for the same two ensembles of 
initial conditions, integrated with $t_{R}=10^{4}t_{D}$.}
\label{landfig}
\end{figure}

\section{THE EFFECTS OF A SUPERMASSIVE BLACK HOLE}

The objective here is to determine how the basic picture described in the
preceding Sections is altered if one allows for a modified potential of the
form
\begin{equation}
V({\bf r})=V_{\gamma}({\bf r})-{M_{BH}\over 
(x^2+y^2+z^2+{\epsilon}^{2})^{1/2}},
\end{equation}
where $V_{\gamma}$ denotes the unperturbed potential associated with the 
density distribution (1). $M_{BH}$ is the black hole mass, expressed in units
of the total galactic mass (since eq. (1) implies that 
$\int\,d^{3}r\,{\rho}({\bf r})=1$) and ${\epsilon}=10^{-3}$ is a softening 
parameter. 
The choice of a relatively large value for ${\epsilon}$ was motivated by 
computational considerations. Noisy simulations are effected with a fixed time 
step integrator (cf. Honerkamp 1994), but integrations with a force which 
becomes very large for 
$r\to 0$ require very small time steps to maintain reasonable accuracy. 
Indeed, even for this comparatively large value of ${\epsilon}$, the black 
hole simulations with $M_{BH}=10^{-3}$ and ${\gamma}=1$ required a time step 
an order of magnitude smaller than for comparable simulations with $M_{BH}=0$.
An investigation of the effects of varying ${\epsilon}$ in the simpler
potential (11), which exhibits qualitatively similar behaviour, indicates that 
the precise value of ${\epsilon}$ is comparatively unimportant, provided only 
that $r_{min}$, the closest approach of an orbit to the black hole, is large 
compared with ${\epsilon}$. This condition suggests that it suffices to 
consider ${\epsilon}{\;}{\ll}{\;}10^{-2}$.

Four sorts of experiments were performed, namely: (1) computations of
``libraries'' of $1000$ representative orbits of length $t=200t_{D}$, to 
determine the relative numbers of regular and chaotic orbits; (2) long time 
integrations for $t=102400t_{D}$ to obtain estimates of the true ${\chi}$ for
chaotic orbits, as well as to extract distributions of short time Lyapunov 
exponents; (3) $1600$-orbit simulations of chaotic mixing, allowing for a 
black hole but no additional perturbations; and (4) $1600$-orbit simulations 
of chaotic mixing that included both a black hole and low amplitude 
time-dependent irregularities. Attention focused primarily on two choices of 
black hole mass, namely $M_{BH}=10^{-3}$ and $M_{BH}=10^{-2}$.

\subsection{How chaotic are the orbits?}

Consistent with Merritt (1998), it was found that allowing for a nonzero
$M_{BH}$ makes the flow `more chaotic' in the sense that, for chaotic orbits,
the typical values of the Lyapunov exponents ${\chi}$ increase. Not 
surprisingly, this effect is most pronounced in the lower energy shells where 
the orbits tend to pass the closest to the hole. For all four values of 
${\gamma}$, introducing a black hole with a mass as large as 
$M_{BH}=10^{-2}$ only increases the typical values of ${\chi}$ in the ninth 
shell by less than a factor of two. However, especially for smaller values 
of ${\gamma}$, the black hole can have much larger effects on the lowest two 
energy shells. For ${\gamma}=0$, a black hole with $M_{BH}=10^{-2}$ increases 
the values of ${\chi}$ in the lowest energy shell by an order of magnitude.
For ${\gamma}=1$, the increase is by a factor of three. For ${\gamma}=2$, a 
black hole with mass $M_{BH}=10^{-2}$ increases the typical ${\chi}$ by less 
than a factor of two, and $M_{BH}=10^{-3}$ has a comparatively minimal effect. 

\begin{figure}
\centering
\centerline{
        \epsfxsize=8cm
        \epsffile{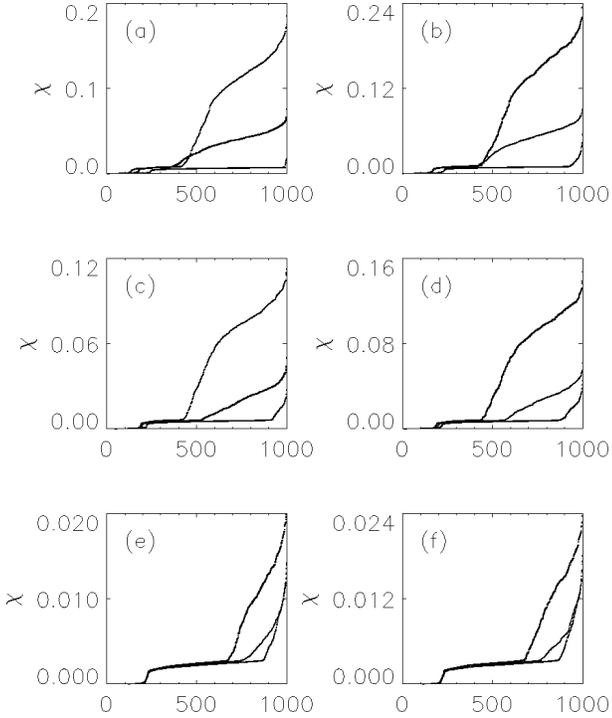}
           }
        \begin{minipage}{9cm}
        \end{minipage}
        \vskip -0.3in\hskip -0.0in
\caption{
(a) Short time ${\chi}(t=200t_{D})$ for $1000$ orbits in the lowest energy 
shell with ${\gamma}=0$, arranged by increasing values of ${\chi}$. From 
bottom to top, the three curves
represent $M_{BH}=0$, $M_{BH}=10^{-3}$, and $M_{BH}=10^{-2}$. (b) The same for
the lowest energy shell with ${\gamma}=0.3$. 
(c) The second lowest energy shell with ${\gamma}=0$.
(d) The second lowest energy shell with ${\gamma}=0.3$.
(e) The ninth lowest energy shell with ${\gamma}=0$.
(f) The ninth lowest energy shell with ${\gamma}=0.3$.}
\label{landfig}
\end{figure}

The presence of a supermassive black hole can also impact significantly the 
relative abundance of ``strongly chaotic'' orbit segments with ${\chi}$ 
significantly larger than zero. For the two less cuspy models, ${\gamma}=0$ 
and ${\gamma}=0.3$, even a black hole mass as small as $M_{BH}=10^{-3}$ 
suffices to trigger a significant increase in the relative number of
strongly chaotic segments, particularly for the lower energy shells. For 
${\gamma}=1$ and ${\Delta}t=200t_{D}$, such a black hole increases 
significantly the number of chaotic orbits in the two lowest energy shells, 
but its effect on the ninth energy shell is much less pronounced. For 
${\gamma}=2$, even a black hole as large as $M_{BH}=10^{-2}$ seems to have no 
appreciable effect on the relative number of strongly chaotic orbits in any
of the three energy shells.

The net conclusion, in agreement with Merritt (1998, 1999), is that, under 
appropriate circumstances, supermassive black holes with $M_{BH}{\;}{\sim}
{\;}10^{-3}-10^{-2}$ can significantly impact both the relative number of 
chaotic orbits that exist and the typical values of their largest short
time Lyapunov exponent. Merritt's argument that the principal effect of the
supermassive black hole is to enhance the strength of the central cusp would
seem to explain why, for ${\gamma}=2$, a black hole of fixed mass has a much
smaller effect than for smaller values of ${\gamma}$: In this case, a very
pronounced density cusp is already there!

But what about the question of `stickiness'? Does the presence of a 
supermassive black hole tend to make distinctions between regular and
chaotic orbits more evident, or is it still possible for a chaotic orbit to
`look regular' for very long times? Here the first obvious point is that,
viewed over times ${\sim}{\;}100t_{D}-200t_{D}$, the computed values of 
${\chi}$ for an ensemble containing both regular and chaotic orbits do not
divide neatly into two disjoint classes with small and large values of ${\chi}$
and nothing in between. 

This is, e.g., manifest from FIGURES 13 and 14, which
summarise information about the numbers of orbit segments with different 
values of ${\chi}$ for integrations of duration $t=200t_{D}$. These FIGURES 
encapsulate information about the short time Lyapunov exponents computed for 
$36$ different ensembles of $1000$ initial conditions -- one each for the 
first, second, and ninth energy shells for the four different values of 
${\gamma}$ with $M_{BH}=0$, $M_{BH}=10^{-3}$, and $M_{BH}=10^{-2}$. For each
ensemble, the orbits were ordered by the size of the computed value of 
${\chi}$ and the resulting ${\chi}$'s were then plotted as a function of 
particle number from $1$ to $1000$. Inspection of these FIGURES thus allows 
one to infer the overall range of ${\chi}$'s that was observed, the relative 
number of orbits in any given interval ${\Delta}{\chi}$, and how this number 
varies with changes in $E$, $M_{BH}$, and/or ${\gamma}$. 
In interpreting these FIGURES, it should be noted that they were generated as 
collections of tiny diamonds, {\it not} solid lines. The near-complete absence 
of any gaps thus manifests the fact that the orbits seemingly sample a 
near-continuous range of ${\chi}$'s, and that there is no pronounced break 
between regular and chaotic orbits. The presumption is that all points 
not lying along a nearly horizontal line with very small ${\chi}$ correspond 
to chaotic orbits, some relatively sticky and some otherwise. 

\begin{figure}
\centering
\centerline{
        \epsfxsize=8cm
        \epsffile{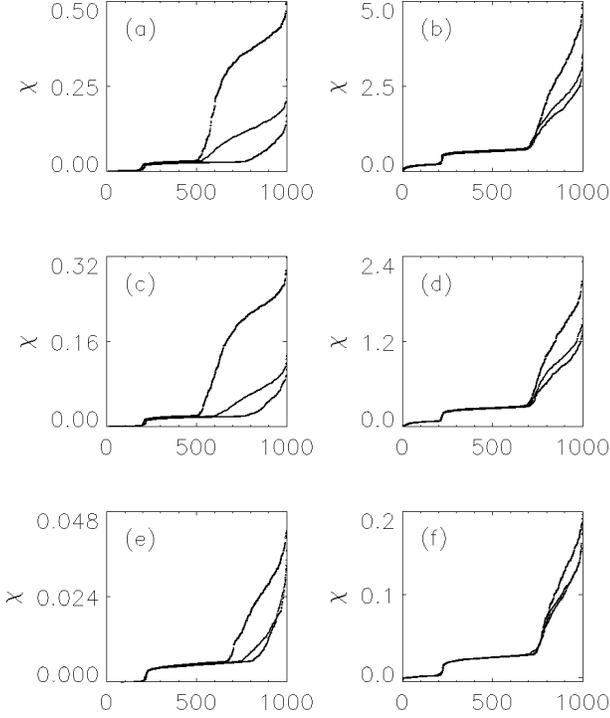}
           }
        \begin{minipage}{10cm}
        \end{minipage}
        \vskip -0.3in\hskip -0.0in
\caption{
(a) Short time ${\chi}(t=200t_{D})$ for $1000$ orbits in the
lowest energy shell with ${\gamma}=1$. From bottom to top, the three curves
represent $M_{BH}=0$, $M_{BH}=10^{-3}$, and $M_{BH}=10^{-2}$. (b) The same for
the lowest energy shell with ${\gamma}=2$. 
(c) The second lowest energy shell with ${\gamma}=1$.
(d) The second lowest energy shell with ${\gamma}=2$.
(e) The ninth lowest energy shell with ${\gamma}=1$.
(f) The ninth lowest energy shell with ${\gamma}=2$.}
\label{landfig}
\end{figure}

The absence of a pronounced gap in ${\chi}$ between the regular and 
chaotic orbits indicates that, as for the case with $M_{BH}=0$, the 
distribution of short time ${\chi}$'s must extend down to zero and, as such,
one might perhaps suppose that $N[{\chi}]$, the distribution of short time
Lyapunov exponents for chaotic orbit segments, is again at least bimodal. That 
this is 
often so is illustrated in FIGURES 15 and 16, which exhibit the analogues of 
several panels from FIGURES 1 and 2 for long time integrations with  
$M_{BH}=10^{-3}$ and $M_{BH}=10^{-2}$.

\begin{figure}
\centering
\centerline{
        \epsfxsize=8cm
        \epsffile{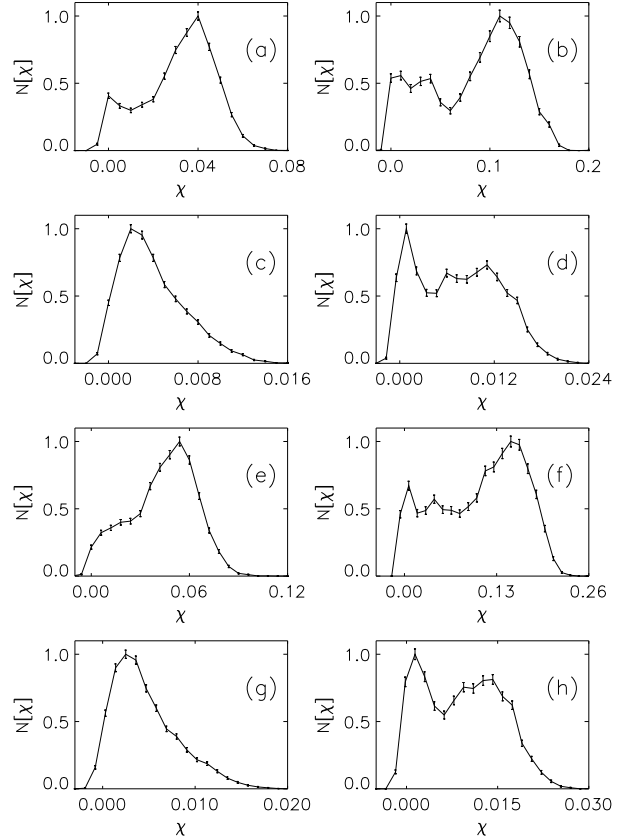}
           }
        \begin{minipage}{10cm}
        \end{minipage}
        \vskip -0.3in\hskip -0.0in
\caption{
(a) $N[{\chi}({\Delta}t)]$, the distribution of short time 
Lyapunov exponents, for chaotic orbits in the lowest energy shell with 
${\gamma}=0$ and $M_{BH}=10^{-3}$ for sampling time ${\Delta}t=100t_{D}$. 
(b) $N[{\chi}({\Delta})]$ for chaotic orbits in the lowest energy shell with 
${\gamma}=0$, $M_{BH}=10^{-2}$, and ${\Delta}t=100t_{D}$.
(c) The same for chaotic orbits in the ninth energy shell with ${\gamma}=0$
and $M_{BH}=10^{-3}$
(d) The same for chaotic orbits in the ninth energy shell with ${\gamma}=0$
and $M_{BH}=10^{-2}$
(e) - (h) The same as (a) - (d) for ${\gamma}=0.3$.}
\label{landfig}
\end{figure}

\begin{figure}
\centering
\centerline{
        \epsfxsize=8cm
        \epsffile{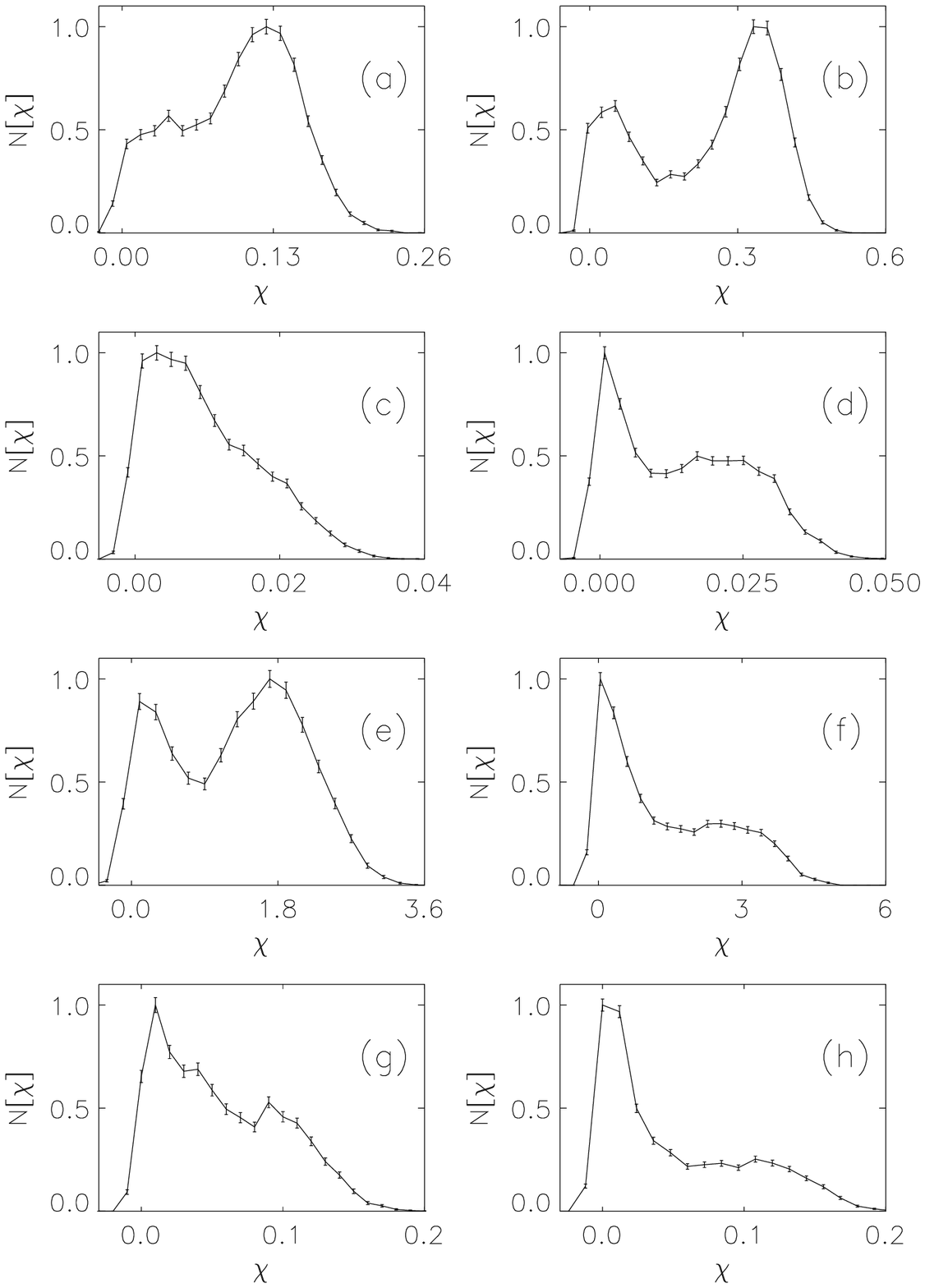}
           }
        \begin{minipage}{10cm}
        \end{minipage}
        \vskip -0.3in\hskip -0.0in
\caption{
(a) $N[{\chi}({\Delta}t)]$, the distribution of short time 
Lyapunov exponents, for chaotic orbits in the lowest energy shell with 
${\gamma}=1$ and $M_{BH}=10^{-3}$ for sampling time ${\Delta}t=100t_{D}$. 
(b) $N[{\chi}({\Delta})]$ for chaotic orbits in the lowest energy shell with 
${\gamma}=1$, $M_{BH}=10^{-2}$, and ${\Delta}t=100t_{D}$.
(c) The same for chaotic orbits in the ninth energy shell with ${\gamma}=1$
and $M_{BH}=10^{-3}$
(d) The same for chaotic orbits in the ninth energy shell with ${\gamma}=1$
and $M_{BH}=10^{-2}$
(e) - (h) The same as (a) - (d) for ${\gamma}=2$.}
\label{landfig}
\end{figure}

But how do things vary with the sampling time ${\Delta}t$? Does the presence
of a supermassive black hole serve to accelerate phase space transport 
significantly, thus making the time scale to diffuse from one portion of phase
space to another much shorter? The answer here seems to be: no! One again 
discovers that, over very long time scales, $t{\;}{\sim}{\;}20000t_{D}$ or 
longer, different
segments of the same chaotic orbit can look extremely different, varying in
appearance from wildly chaotic to nearly regular. This is manifested by the
time-dependence of the dispersions ${\sigma}_{\chi}({\Delta}t)$ which, as for
the case when $M_{BH}=0$, typically decay much more slowly than as 
${\Delta}t^{-1/2}$. Examples of the observed behaviour, generated as for 
FIGURES 3 and 4, are exhibited in FIGURE 17, which show ${\sigma}_{\chi}$ for
several different choices of ${\gamma}$, $E$, and $M_{BH}$.

\begin{figure}
\centering
\centerline{
        \epsfxsize=8cm
        \epsffile{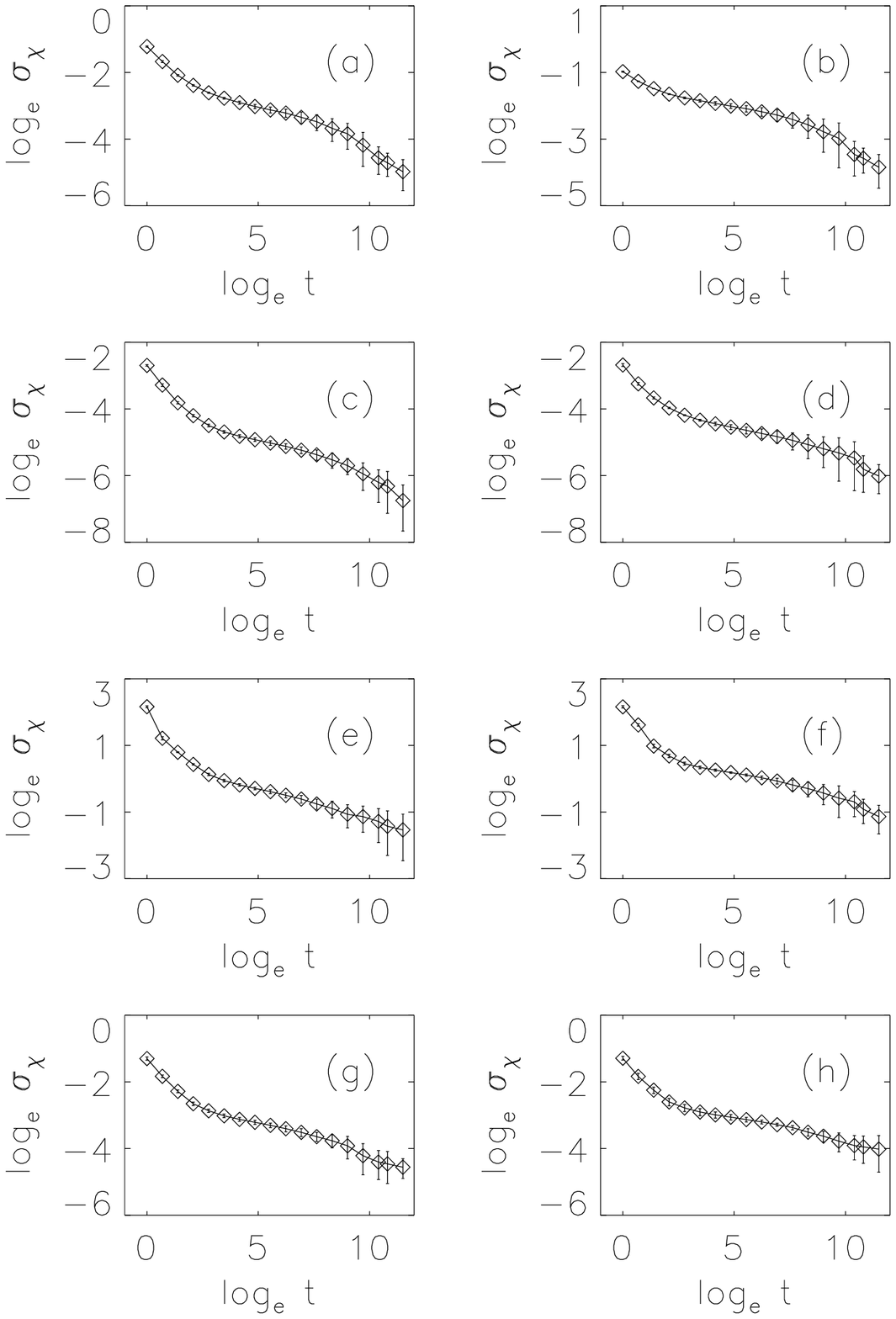}
           }
        \begin{minipage}{10cm}
        \end{minipage}
        \vskip -0.3in\hskip -0.0in
\caption{
(a) The dispersion ${\sigma}_{\chi}$ for chaotic orbits in 
the lowest energy shell with ${\gamma}=1$ and $M_{BH}=10^{-3}$, computed as a 
function of sampling time ${\Delta}t$. 
(b) ${\sigma}_{\chi}$ for chaotic orbits in the lowest energy shell with 
${\gamma}=1$ and $M_{BH}=10^{-2}$. 
(c) The same for the ninth energy shell and $M_{BH}=10^{-3}$.
(d) The same for the ninth energy shell and $M_{BH}=10^{-2}$.
(e) - (h) The same as (a) - (d) for ${\gamma}=2$.}
\vspace{0.0cm}
\end{figure}

\begin{figure}
\centering
\centerline{
        \epsfxsize=8cm
        \epsffile{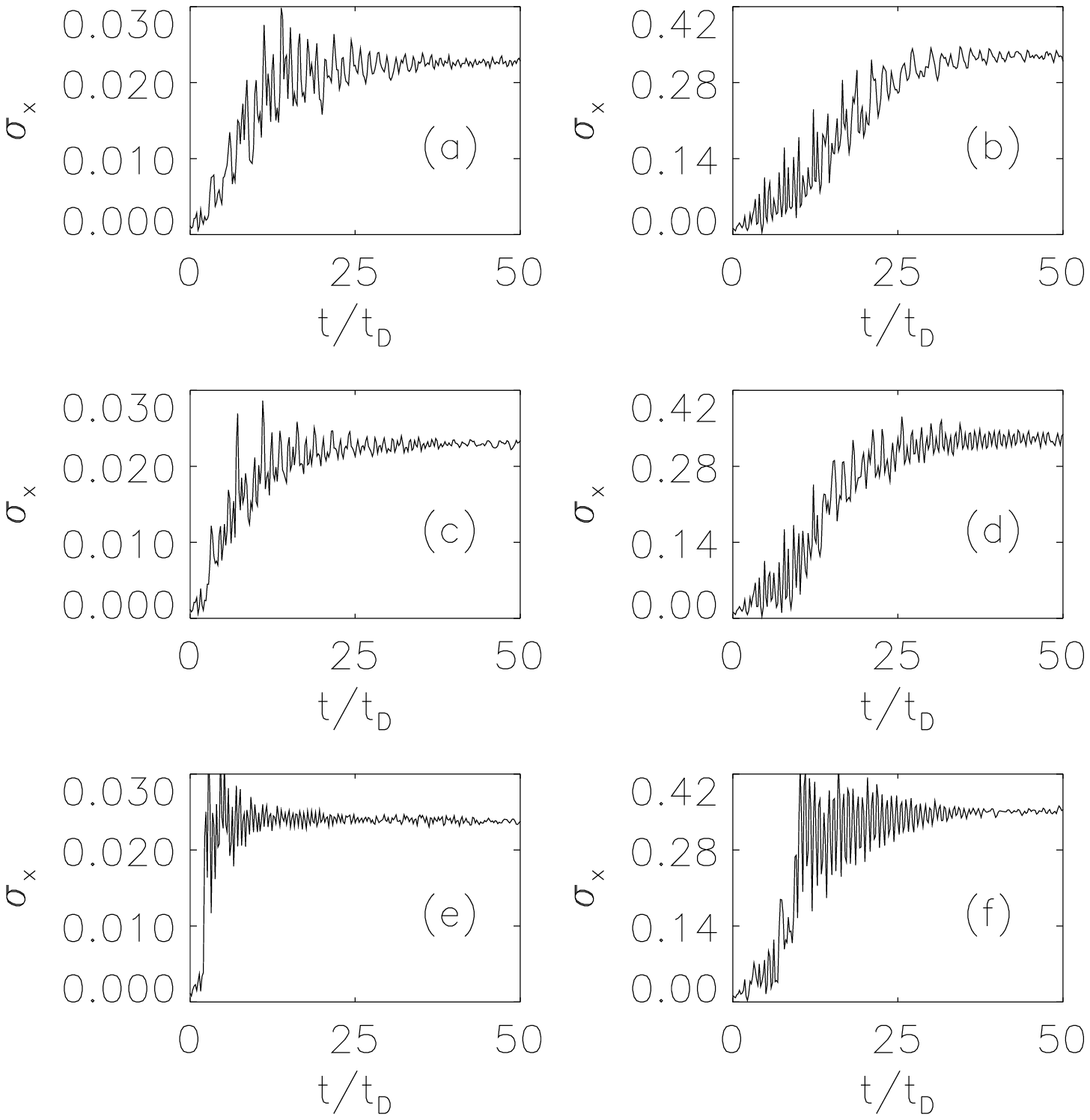}
           }
        \begin{minipage}{10cm}
        \end{minipage}
        \vskip -1.45in\hskip -0.0in
\caption{
(a) The dispersion ${\sigma}_{x}$ computed for an 
unperturbed ensemble of $1600$ orbits in the second lowest energy shell for 
${\gamma}=2$, evolved in the absence of a supermassive black hole.
(b) The same for an ensemble in the ninth energy shell. (c) The same as (a),
but now allowing for a black hole with mass $M_{BH}=10^{-3}$. (d) The same
as (b), but with $M_{BH}=10^{-3}$. 
(e) The same as (c), but with $M_{BH}=10^{-2}$. 
(f) The same as (d), but with $M_{BH}=10^{-2}$. }
\label{landfig}
\end{figure}
\subsection{Chaotic mixing}

The effects of a supermassive black hole on chaotic mixing are completely
predictable. The presence of such a black hole fails to facilitate
significantly the evolution of initially localised ensembles towards a true 
equilibrium in the sense that the phase space is still very sticky, so that
chaotic orbits can remain confined in restricted phase space regions for very
long times. However, a supermassive black hole {\it does} tend to accelerate
the approach towards a near-equilibrium. The correlation between the Lyapunov
time $t_{L}=1/{\chi}$ and the time scale $t_{M}$ on which a chaotic ensemble 
evolves persists for $M_{BH}{\;}{\ne}{\;}0$, but this means that larger 
${\chi}$'s correspond to a more rapid evolution. One example of this
behaviour is shown in FIGURE 18, which exhibits ${\sigma}_{x}$ for the same
initial ensembles evolved with ${\gamma}=2$ and variable $M_{BH}=0$, 
$M_{BH}=10^{-3}$, and $M_{BH}=10^{-2}$.

The effects of weak perturbations on chaotic mixing with a 
supermassive black hole are also predictable. As is the case for $M_{BH}=0$,
such perturbations can facilitate phase space transport, thus accelerating
the approach towards equilibrium; and, even in the absence of any obvious
`stickiness,' they can play an important role in `fuzzing out' high 
frequency structures. Moreover, the minimum amplitudes required to have an
appreciable effect are comparable to what is required when $M_{BH}=0$. In
particular, discreteness effects, modeled as friction and white noise, should
be important on a time scale $<100t_{D}$ provided only that $t_{R}<10^{6}-
10^{7}t_{D}$.

\section{DISCUSSION}

The `stickiness' observed in these triaxial Dehnen potentials is so striking
that one may wonder how generic it is. It is thus reassuring that 
the same qualitative behaviour can arise for the simple model comprised of an
anisotropic oscillator and a spherical Plummer potential. For example, the
qualitative form of $N[{\chi}({\Delta}t)]$, the distribution of short time
Lyapunov exponents for the triaxial Dehnen potentials, can be reproduced by 
orbits evolved in a potential of the form
\begin{equation}
V({\bf r})={1\over 2}{\Bigl(}{\omega}_{x}^{2}x^{2}+{\omega}_{y}^{2}y^{2}+
{\omega}_{z}^{2}z^{2}{\Bigr)}-
{M_{BH}\over \sqrt{r^{2}+{\epsilon}^{2}}},
\end{equation}
with ${\omega}_{x}^{2}=1.0$, ${\omega}_{y}^{2}=1.25$, ${\omega}_{z}^{2}=0.75$, 
and ${\epsilon}=10^{-3}$, for appropriate choices of energy $E$ and black hole
mass $M_{BH}$. 

For $M_{BH}=0$ and $M_{BH}\to\infty$, motion in this potential is completely
integrable. However, $V({\bf r})$ is nonintegrable for intermediate values
and, for a reasonable range of black hole masses, admits a coexistence of 
large numbers of both
regular and chaotic orbits. In this range, one finds typically that increasing
$M_{BH}$ makes the system more chaotic in the sense that the values of 
${\chi}$ tend to increase, but that, when expressed in units of $1/t_{D}$,
the changes in ${\chi}$ are not all that large. One discovers further that,
generically, the chaotic orbits tend to be very sticky, so that an integration
for a time $t=10000t_{D}$ or longer does not suffice to yield convergence 
towards the asymptotic ${\chi}$, defined in a $t\to\infty$ limit. In both
these ways, this potential is very Dehnenesque.

As an indication of the degree to which this potential yields similar 
distributions of short time Lyapunov exponents $N[{\chi}({\Delta}t)]$, 
consider 
FIGURE 19, which exhibits $N[{\chi}({\Delta}t)]$ for ensembles of chaotic 
orbits. These ensembles were generated from initial conditions that uniformly 
sampled the $y-z$ plane, setting $x=v_{y}=v_{z}=0$ and, for given energy $E$, 
solving for $v_{x}(y,z,E)>0$. Each orbit was integrated for a time $t=4096$ 
and the computed values of ${\chi}$ at that time were used to identify which 
initial conditions appeared chaotic. The orbits identified as chaotic were 
then 
divided into $16$ segments of length ${\Delta}t=256$ which, for the range of 
energies and black hole masses exhibited in these FIGURES, corresponded to a 
period of order $100t_{D}$, and short time Lyapunov exponents were identified 
for each segment. The resulting collection of ${\chi}$'s was then binned to
generate $N[{\chi}]$ for ${\Delta}t=256$. The first panel, for 
$M_{BH}=10^{-1.5}$ and $E=0.0044$, closely resembles the bimodal distributions 
$N[{\chi}({\Delta}t)]$ observed for the lower energy shells in the Dehnen 
potentials. The second panel, generated for $M_{BH}=10^{-2}$ and $E=0.65$, 
resembles more closely the distributions for the higher energy shells, 
especially in the presence of a black hole.

The results described in this paper have potentially significant implications 
for modeling galaxies using Schwarzschild's (1979) method or any comparable
technique. One obvious point is that, in the absence of any time-dependent
perturbations, phase space transport can be {\it extremely} inefficient, so
that computing an ensemble of orbits even for $1000t_{D}$ or longer may not
suffice to obtain a reasonable approximation to an invariant distribution.
Great care needs to be taken in endeavoring to create truly time-independent
building blocks. At least for cusps that are not too steep, inserting a 
supermassive black hole typically makes orbits more chaotic in the sense that
the Lyapunov exponents become larger and, in addition, tends to increase the
relative abundance of chaotic orbits. However, the presence of the black hole
does {\it not} seem to significantly accelerate the rate of phase space 
transport.

\begin{figure}
\centering
\centerline{
        \epsfxsize=8cm
        \epsffile{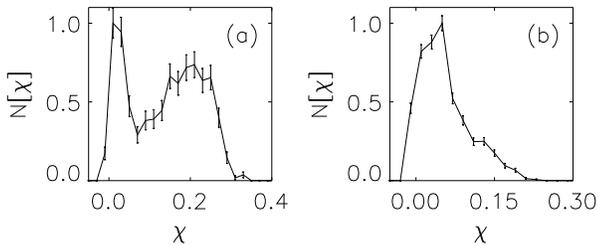}
           }
        \begin{minipage}{10cm}
        \end{minipage}
        \vskip -3.0in\hskip -0.0in
\caption{
(a) Short time $N[{\chi}(t=256)]$ for a collection of 
chaotic orbits evolved in the toy potential (11) with $M_{BH}=10^{-1.5}$ and 
$E=0.044$. (b) The same for $M_{BH}=10^{-2}$ and $E=0.065$.}
\label{landfig}
\end{figure}

The other obvious point is that even comparatively weak time-dependent 
perturbations tend to significantly accelerate phase space diffusion, allowing
orbits to probe different phase space regions on times short compared with
the time scale for phase space transport in the absence of any perturbations.
This basic fact, first recognised in the context of simple maps (cf. Lieberman
\& Lichtenberg 1972) and subsequently explored for simple two- and 
three-degree-of-freedom systems (cf. Habib, Kandrup, \& Mahon 1997), persists
for seemingly realistic cuspy, triaxial potentials. This suggests immediately
that irregularities associated with the real world could serve to destabilise
pseudo-equilibria constructed using near-invariant, rather than truly 
invariant, building blocks, even if, in the absence of perturbations, such
equilibria could persist for times ${\gg}{\;}t_{H}$.

But what are the effects of perturbations like friction and noise? Do they
simply accelerate the evolution of an unperturbed system or do they alter the
final state? Noise induces significant changes in the energy and any other 
global isolating integrals on a relaxation time $t_{R}=1/{\eta}$ and, as 
such, it is clear that noise must change the form of the final state for later
times. However, there is good reason to anticipate that, for times 
$t{\;}{\ll}{\;}t_{R}$, noise simply speeds things up without significantly
altering the 
ultimate outcome. In other words, an ensemble evolved for a time $t_{H}$ in
the presence of weak noise with $t_{R}{\;}{\gg}{\;}t_{H}$ might be expected to 
achieve a final state closely resembling the state which an unperturbed 
ensemble would have achieved only after a time 
$100t_{H}$, or even longer. The basic idea is that the perturbations drive 
phase space transport by helping stars to find phase space holes, thus 
accelerating the approach towards an equilibrium, but that, at least to the
extent that the perturbations do not significantly alter the values of the
energy and other isolating integrals, they will not impact the form of the
final equilibrium.

An analogy with kinetic theory were perhaps appropriate. When analysing 
interactions of particles in a dilute gas, one speaks of microreversibility,
which implies that the transition rates 
$W({\bf v}_{1},{\bf v}_{2}\to {\bf v}_{3},{\bf v}_{4})$ and
$W({\bf v}_{3},{\bf v}_{4}\to {\bf v}_{1},{\bf v}_{2})$ must be equal.
This is a statement about the phase space kinematics, and is without reference
to the dynamics. Statistical equilibrium requires detailed balance, 
characterised by populations $f({\bf v}_{i})$ satisfying 
$f({\bf v}_{1})f({\bf v}_{2})
W({\bf v}_{1},{\bf v}_{2}\to {\bf v}_{3},{\bf v}_{4})=
f({\bf v}_{3})f({\bf v}_{4})
W({\bf v}_{3},{\bf v}_{4}\to {\bf v}_{1},{\bf v}_{2}).$ 

In the context of
phase space diffusion, the analogue of microreversibility is the notion that
the transition rates $W(A\to B)$ and $W(B\to A)$ across a leaky barrier
separating regions $A$ and $B$ must be equal. Equilibrium involves a detailed
balance with equal (and constant) number densities on both sides of the 
boundary, so that the product of number density and transition rate is the
same. That $W(A\to B)=W(B\to A)$ is incorporated explicitly in every theory
of phase space transport, including, e.g., the ``turnstile model'' for 
diffusion through cantori, which is the most successful to date (MacKay, Meiss,
\& Percival 1984). The expectation, then, is that low amplitude perturbations
will help stars breach these leaky barriers, but that they will not alter the
balance associated with an equilibrium. One could, perhaps, envision carefully
tailored ``designer noise'' so constructed as to favour motion in one direction
over the other, but this would seem contrived. The noise used in this paper,
and most other investigations of accelerated phase space transport, was chosen
to act equally on stars on both sides of the boundary; and, as such, should
not serve to alter the final balance.

There is also concrete numerical
evidence (cf. Habib, Kandrup, \& Mahon 1996, 1997) suggesting that weak noise 
simply accelerates phase space transport rather than leading to a new 
equilibrium. This derives from comparing the same ensemble of initial 
conditions evolved in three different ways, viz: (i) for comparatively short 
times, $t{\;}{\le}{\;}100t_{D}$, in the absence of any perturbations, (ii) for 
longer times, $t>1000t_{D}$, in the absence of perturbations, and (iii) for 
short times, again $t{\;}{\le}{\;}100t_{D}$, in the presence of weak noise
for which $t_{R}{\;}{\gg}{\;}100t_{D}$. For some choices of potential and/or
ensemble, one finds that the short and long time unperturbed simulations 
result in very similar end states, as probed, e.g., by lower order moments
and coarse-grained distribution functions. In this case, 
weak noise has only a minimal effect. In other cases, however, the short
and long times unperturbed simulations yield substantially different end 
states. In these cases, one finds invariably that the noisy simulations yield
a final state that more closely resembles the late time unperturbed state than
did the shorter time unperturbed evolution. Moreover, for noise of sufficient
strength, one discovers oftentimes that the noisy end state and the long time
unperturbed endstate are almost identical.

This motivates a specific proposal, namely that, when using 
Schwarzschild's method to construct models that contain significant numbers
of chaotic orbits, one should generate ``noisy libraries,'' which are more
likely to constitute reasonable approximations to invariant distributions 
than are libraries generated without any perturbations. There is no guarantee
that the noise will yield truly time-independent building blocks. However, it
seems reasonable to anticipate that allowing for weak noise will yield 
libraries that are, at least, better approximations to invariant distributions.
\vskip .2in
In contemplating the implications of the computations described in this paper
for real galaxies, it is important to recall that the notion of an equilibrium
is an idealisation which is never achieved in the real world. Assuming that
the evolution is not strongly dissipative, so that Liouville's Theorem holds
at least approximately, there can at best be a coarse-grained approach towards
an equilibrium (cf. Lynden-Bell 1967). One might anticipate an approach towards
equilibrium in the sense that various moments and/or coarse-grained 
distributions become systematically more nearly time-independent, but there
can be no true pointwise approach towards equilibrium. Moreover, realistic
systems are continually subjected to various time-dependent perturbations
reflecting both internal and external irregularities, so that, even allowing
for dissipation, one could at best view a galaxy as being ``near'' equilibrium.
The basic question is whether it be reasonable to assume that a real galaxy
evolves towards a complex time-dependent state which can be approximated 
realistically as an equilibrium.

Evolving towards a true triaxial equilibrium might seem relatively difficult!
It would seem unreasonable to model realistic triaxial systems as equilibria
that depend only on one global integral and nothing else. A one-integral $f(E)$
depending only on energy $E$ must be spherical (cf. Perez \& Aly 1996). It
{\it is} possible to construct rotating triaxial equilibria $f(E_{J})$ which 
involve only the Jacobi integral $E_{J}$, but such equilibria apparently 
cannot be sufficiently centrally condensed to mimic real physical systems (cf. 
Vandervoort 1980, Ipser \& Managan 1981). Potentials that admit two
or three global integrals are a set of measure zero in the set of 
three-dimensional potentials, and there is a precise sense in which 
near-integrable systems are substantially rarer for three- and 
higher-degree-of-freedom systems than for systems with two degrees of freedom
(Poincar\'e 1892).

There remains the possibility of trying to construct equilibria involving
non-standard local integrals in addition to the conventional global isolating
integrals, perhaps excluding the chaotic orbits altogether. This 
possibility seems implicit in virtually all work on
cuspy triaxial systems hitherto (cf. Zhao 1996, Siopis 1998) and an attempt 
has been made recently to do this in a more systematic fashion (H\"afner et 
al 1999). Most, if not all, of the potentials that 
have been considered, including the triaxial Dehnen potentials, have
chaotic orbits with two positive Lyapunov exponents and, consequently, can
admit only one global isolating integral. However, many dynamicists seem 
to have the intuitive expectation that such equilibria would be hard to 
realise, since they entail a ``fine-tuning'' on a comparatively microscopic
level. As such, one might therefore ask: is it reasonable to expect that the
true time-dependent distribution function associated with a real galaxy 
actually manages to approach one of these finely tuned distributions involving
local integrals?

Irrespective of the answer to this question, it would seem substantially
easier to approach a near-equilibrium than a true equilibrium. One would
anticipate intuitively that there are many more near-equilibria than true
equilibria, and these could be supported approximately with 
two types of building blocks, namely (i) building blocks that are only
approximately time-independent and/or (ii) building blocks which, albeit 
exactly time-independent, only reproduce the potential approximately.
In this regard, it should perhaps be stressed that there is no guarantee 
that, in the absence of dissipation, a real galaxy will evolve towards a
time-independent equilibrium. One could, at least in principle, envision 
an evolution towards a final state that entails undamped, finite amplitude 
oscillations about some time-independent equilibrium (cf. Louis \& Gerhart 
1988, Sridhar 1989).

These observations are related to two significant limitations implicit in
Schwarzschild's method or any obvious variant thereof: (1) There is
no guarantee that any solution constructed using Schwarzschild's method
corresponds to a true equilibrium. It is possible that, strictly speaking, the
assumed potential admits no true equilibria, and that the numerical algorithm
has only constructed an approximate equilibrium (or worse). (2) The fact
that, for some assumed potential, one's orbit library could not be used to
generate a Schwarzschild equilibrium does not imply that an equilibrium
does not exist; and even if there is no equilibrium this does not preclude
the possibility that there exist ``nearby'' potentials that correspond to a
system which (at least on the average) only evolves comparatively slowly.

The enormous stickiness manifested by chaotic orbits in (at least some) cuspy
triaxial potentials can facilitate the construction of approximate equilibria.
However, there is every reason to think that low amplitude perturbations 
associated with both internal and external irregularities could trigger 
nontrivial evolutionary effects on a time scale ${\ll}{\;}t_{H}$.

From this slightly different chain of reasoning, one is led to a conclusion
similar to that of Merritt (1996), namely that galaxies could exhibit a 
two-stage evolution, a rapid approach towards a roughly, but not truly, 
time-independent equilibrium on a time scale ${\sim}{\;}t_{D}$ which is
succeeded by a substantially slower evolution as the system ``searches'' for
a true equilibrium. Using this line of reasoning exclusively, there is no
obvious guarantee that this slower evolution will lead to configurations more
nearly axisymmetric -- one might instead see an evolution through a sequence
of comparably triaxial equilibria. However, the apparent fact that very
cuspy galaxies tend to be more nearly axisymmetric, at least in the center,
could reflect in part the fact that, in these high density central regions,
dissipation was sufficiently strong to allow the system to ``develop a global
conserved quantity.''

\section*{Acknowledgments}
CS thanks the Observatoire de Marseille for generous hospitality during the
later stages of this project. Portions of the manuscript were written while 
HEK was a visitor at the Aspen Center for Physics, the hospitality of which is 
acknowledged gratefully. Thanks also to Brent Nelson, the Florida Astronomy 
Systems Manager, for assistance in coordinating the use of the 24 {\it CPU}'s 
used for the integrations reported here.

\end{document}